\documentclass[%
 reprint,
 amsmath,amssymb,
 aps,
]{revtex4-2}

\usepackage[dvipsnames]{xcolor}
\usepackage{graphicx}
\usepackage{dcolumn}
\usepackage{threeparttable}

\usepackage{booktabs} 
\usepackage{tabularx}
\usepackage[normalem]{ulem} 

\usepackage{xeCJK}

\usepackage{amsmath}
\usepackage{amssymb}
\newcommand{\aap}{Astronomy \& Astrophysics}

\usepackage{bm}

\usepackage{aas_marcos}

\begin{document}
\preprint{APS/123-QED}

\title{Collective Winds of Massive Star Clusters as the Dominant PeVatrons for Galactic Cosmic Rays}
\author{Zijian Qiu}
\affiliation{School of Physics and Astronomy, Sun Yat-sen University, Zhuhai, 519082, China}

\author{Sujie Lin}
\email{linsj6@mail.sysu.edu.cn}
\affiliation{School of Physics and Astronomy, Sun Yat-sen University, Zhuhai, 519082, China}

\author{Lili Yang}%
\email{yanglli5@mail.sysu.edu.cn}
\affiliation{School of Physics and Astronomy, Sun Yat-sen University, Zhuhai, 519082, China}
\affiliation{Centre for Astro-Particle Physics, University of Johannesburg, PO Box 524, Auckland Park 2006, South Africa}

\date{\today}

\begin{abstract}
The knee feature in the cosmic-ray energy spectrum around 4~PeV is widely believed to have a Galactic origin, but the acceleration mechanism and identification of PeVatrons remain key open questions in high-energy astrophysics. Recent precise measurements by LHAASO reveal that the proton and helium spectra exhibit a common rigidity-dependent spectral break at $\sim$3.5~PV, imposing a stringent constraint on source models. In this work, we construct, for the first time, a time-dependent cosmic-ray injection model that incorporates the full evolution of massive stars together with the dynamical development of wind termination shocks. We find that stellar winds of individual massive stars cannot explain the common spectral break observed by LHAASO, as they yield distinct rigidity cutoffs for protons and helium. By contrast, collective winds of massive star clusters naturally reconcile this discrepancy through the mixing effect of stars at different evolutionary stages. We propose a stellar-dominated model in which supernova remnants dominate the GeV--TeV range, individual stellar winds dominate the TeV range, and collective cluster winds dominate the PeV knee region. This model successfully reproduces the rigidity-dependent spectral features of various species near 100~GV and 0.1~PV. It further makes two testable predictions for future observations. Around 0.5 PV, the energy spectra of carbon and oxygen are expected to exhibit hardening similar to that of helium, which can be verified by LHAASO observations. In the multi-TV range, the energy spectrum of magnesium is not expected to show hardening similar to that observed for helium, carbon, and oxygen, which can be tested by DAMPE observations.

\end{abstract}

\maketitle


\section{Introduction} \label{sec:introduction}

Cosmic rays (CRs) with energies extending to the ``knee'' at approximately 4 PeV are widely believed to originate within the Milky Way.
However, this energy scale exceeds the theoretical maximum energy achievable by the standard diffusive shock acceleration (DSA) in supernova remnants (SNRs), the primary CR accelerators below 100 TeV~\cite{lagageMaximumEnergyCosmic1983}.
This discrepancy has intensely driven the search for PeVatrons, Galactic sources capable of accelerating particles to PeV energies. Recent PeV gamma-ray observations have enabled the robust identification of multiple PeVatron candidates, including peculiar SNRs~\cite{caoUltrahighenergyPhotons142021, amenomoriPotentialPeVatronSupernova2021}, massive stellar clusters (MSCs)~\cite{collaborationUltrahighenergyGammarayBubble2024}, microquasars~\cite{collaborationUltrahighEnergyGammarayEmission2025},
Sagittarius A*~\cite{collaborationAccelerationPetaelectronvoltProtons2016},
and pulsar wind nebulae~\cite{2026NatAs.tmp...79L}. However, the dominant source of Galactic PeV CRs remains elusive, as gamma rays often suffer from leptonic contamination. Although neutrino observations and CR composition measurements have the potential to answer this question, these endeavors require either high statistics or accurate particle identification.

LHAASO has recently transformed this situation. Unlike previous ground-based arrays that relied on statistical unfolding for composition analysis (e.g., KASCADE~\cite{collaborationEnergySpectraElemental2009,apelKASCADEGrandeMeasurementsEnergy2013} and IceTop~\cite{icecubecollaborationCosmicRaySpectrum2019}), LHAASO's hybrid detection technique enables event-by-event composition measurements. These measurements revealed that both proton and helium spectra exhibit spectral breaks at a common rigidity of about 3.5 PV~\cite{collaborationFirstIdentificationPrecise2025,collaborationPreciseMeasurementCosmic2025}. This result is consistent with the standard picture in which both CR acceleration and propagation are rigidity dependent~\cite{Peters:1961mxb, DAMPE:2025opn}. At the same time, it places an exceptionally stringent constraint on PeVatron models: different nuclear species must share a common rigidity cutoff in their injection spectra.


Among the proposed PeVatron candidates, the stellar winds of massive stars have drawn considerable attention, because of their large energy budget and the anomalous $^{22}\mathrm{Ne}/ ^{20}\mathrm{Ne}$ isotopic ratio~\cite{1983SSRv...36..173C,2020SSRv..216...42B,2020MNRAS.493.3159G}. In particular, MSCs can drive powerful collective winds through the superposition of individual stellar outflows, making them promising sites for PeV particle acceleration~\cite{2021MNRAS.504.6096M,2023MNRAS.519..136V}. Gamma-ray observations of regions such as Cyg~OB2~\cite{collaborationUltrahighenergyGammarayBubble2024,2019NatAs...3..561A} and Westerlund~1~\cite{2019NatAs...3..561A}, together with theoretical studies~\cite{2020MNRAS.493.3159G,2021MNRAS.504.6096M}, indicate that the wind termination shocks (WTS) formed by these stellar winds can efficiently accelerate particles to PeV energies through DSA.

Nevertheless, previous studies typically assume a universal injection rigidity spectrum for all species, overlooking the crucial fact that massive stars evolve. During their main-sequence lifetime, massive stars exist as O-type stars, producing relatively slow, proton-rich winds, before transitioning into a Wolf-Rayet (WR) phase, characterized by fast, dense, helium-dominated outflows. These two phases differ markedly in wind power, composition, and shock properties, naturally imprinting distinct rigidity cutoffs on the proton and helium injection spectra. Whether stellar winds remain viable PeVatron candidates once such evolutionary effects are properly accounted for has not been examined. Addressing this question requires an injection model that explicitly incorporates stellar evolution.

In this work, we combine for the first time the full evolutionary sequence of massive stars with the dynamical evolution of WTS to construct a time-dependent injection model. We apply this framework to both individual stellar winds and collective winds of MSCs. Our analysis reveals a clear distinction. Individual stellar winds inevitably produce rigidity cutoffs for proton and helium that deviate significantly from each other and are therefore inconsistent with LHAASO observations. The mixing effect of collective stellar winds, however, naturally reconciles this discrepancy, identifying MSCs as the only viable PeVatron scenario among stellar-wind sources. Motivated by this finding, we propose a self-consistent stellar-dominated model in which SNRs, individual stellar winds, and collective stellar winds respectively dominate distinct energy ranges of the CR spectrum, and we show that this model can explain the rigidity-dependent differences in the energy spectra of various elemental species at 100 GV and 0.1 PV.

The paper is organized as follows. In Section \ref{sec:methods}, we describe the framework for computing time-dependent injection spectra, including stellar evolution modeling, WTS dynamics, and the calculation of maximum rigidities and injection rates. Section \ref{sec:result} compares the model predictions with the LHAASO observations and shows that only the collective-wind scenario remains consistent with the data. In Section \ref{sec:CWM}, we introduce a stellar-dominated model and compare its predictions with the rigidity spectra of various species in the GeV–PeV energy range. Finally, Section \ref{sec:dis} presents the discussion and conclusions.


\section{Evolution-Dependent Cosmic-Ray Injection from Stellar Winds} \label{sec:methods}

\begin{figure}[htpp]
    \centering
    \includegraphics[width=1\linewidth]{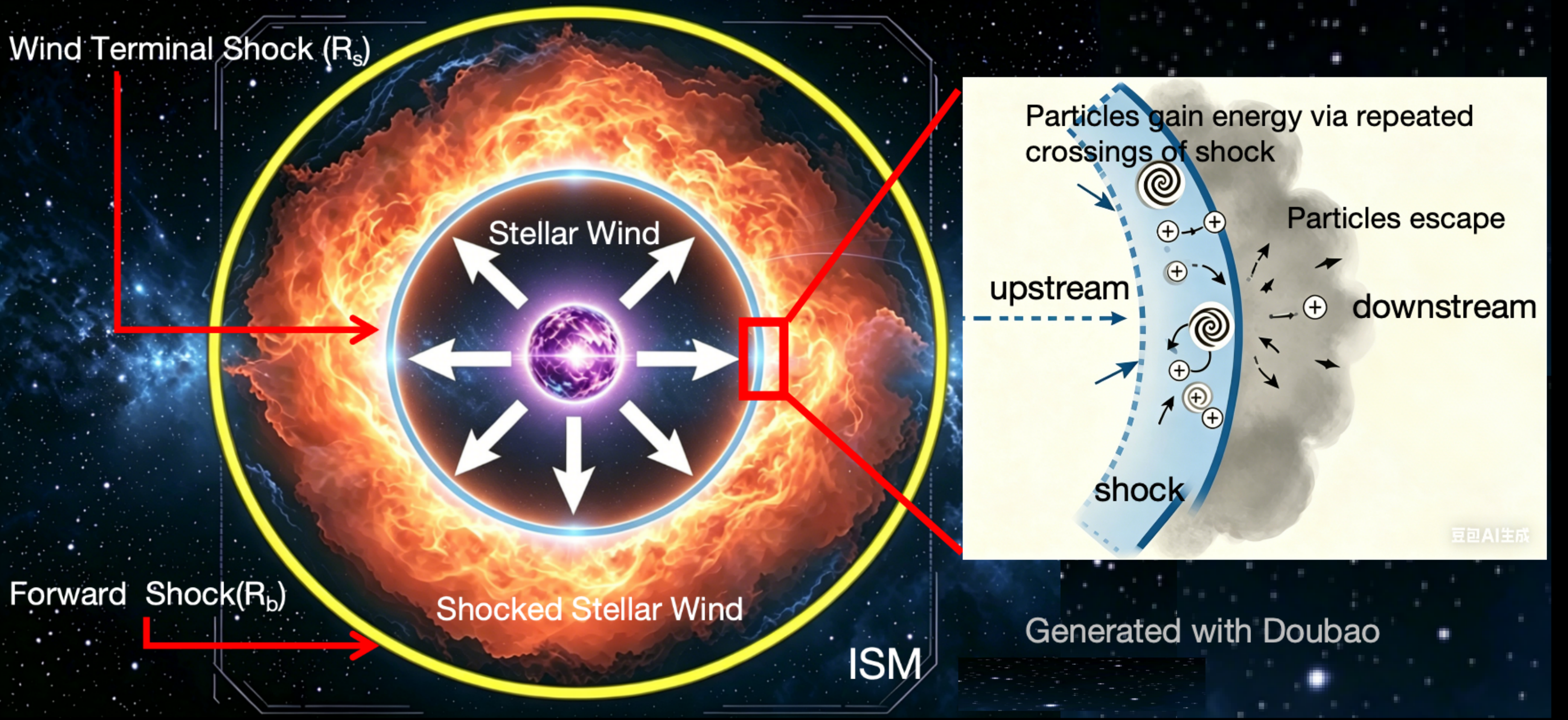}
    \caption{Schematic of stellar WTSs and CR acceleration. For MSCs, dense stellar populations replace the single star shown in the figure and generate outward collective stellar winds.}
    \label{fig:Schematic}
\end{figure}

The stellar winds of massive stars, as they expand into the surrounding interstellar medium (ISM), sweep up ambient material and forms a forward shock at $R_b$. Inside \(R_b\), the shocked stellar wind is fully thermalized, giving rise to a hot bubble known as a wind-driven bubble. Meanwhile, the supersonic stellar wind from the star interacts with thid shocked wind at $R_s$, producing a shock front referred to as the WTS, as shown in Fig.~\ref{fig:Schematic}. The location of the WTS is determined by the balance between the ram pressure of the stellar wind and the thermal pressure within the bubble. This WTS serves as a natural site for DSA, where particles can gain energy by repeatedly crossing the shock front, producing a non-thermal spectrum that extends up to a maximum rigidity set by the age and size of the shock. In dense MSCs, the stellar winds of individual stars can merge with one another, forming a global WTS on scales of tens of parsecs, resulting in substantially greater energy injection than that of individual stellar winds. 
To quantify the injection spectrum of both individual and collective WTS, we study how the wind properties, such as mass-loss rate, composition, and terminal velocity, vary over the evolutionary lifetime of the driving stars and how these variations are imprinted on the spectrum of accelerated particles. 

In the following text, stellar winds refer to the individual winds of massive stars, including both O-type and WR stars, whereas collective winds denote the large-scale outflow of an MSC. Since the masses of MSCs that have been proposed as potential hosts of collective winds generally exceed $10^4 M_{\odot}$ (e.g., Quintuplet $\sim 1\times 10^4 M_{\odot}$~\cite{portegies_zwart_young_2010}, Westerlund 1 $\sim 3.1\times 10^4 M_{\odot}$~\cite{portegies_zwart_young_2010}, NGC 3603 $\sim 1.3\times 10^4 M_{\odot}$~\cite{portegies_zwart_young_2010}), we neglect the density requirements for the formation of a collective wind and assume that any MSC with a total mass above $10^4 M_{\odot}$ can host a collective wind. Furthermore, in all collective-wind calculations presented in this work, we adopt a cluster mass of $3\times 10^4M_{\odot}$, comparable to that of Cyg OB2~\cite{Wright:2010cc} and representative of a typical MSC.

\subsection{Stellar Evolution}\label{subsec:stellar_evolution_modeling}

The elemental abundances in stars evolve through nucleosynthesis and mixing processes such as microscopic diffusion, gravitational settling, convection, and magnetic effects.
These processes are incorporated in rotating, mass-losing stars using the Modules for Experiments in Stellar Astrophysics (MESA) framework.
In this work, we adopt evolutionary tracks and isochrones from the MESA Isochrones and Stellar Tracks (MIST) project~\cite{choiMesaIsochronesStellar2016}, which has integrated updated observational constraints into MESA simulations, to follow the evolution of abundances and mass-loss behavior of stars with initial masses from $8~M_{\odot}$ to $150~M_{\odot}$.

The terminal velocity of the stellar wind, $v_\infty$, is a critical parameter for shock dynamics, but is not provided as a standard output of MESA.
We estimate it using physically motivated prescriptions tailored to each stellar type:

\begin{itemize}
\item O-type stars are defined here as stars with an effective temperature $T_\mathrm{eff} > 1.1\times10^{4}\,\mathrm{K}$ and a surface hydrogen mass fraction $X_H > 0.4$.
Their winds are modeled using the line-driven radiation wind theory of Castor, Abbott \& Klein (CAK) \cite{castorRadiationdrivenWindsStars1975}.
Following \citet{2001A&A...369..574V}, we assume that the terminal velocity scales with the stellar escape velocity $v_{\mathrm{esc}}$ and the surface metallicity $Z_{\mathrm{surf}}$ as
\begin{equation}
v_\infty =
\begin{cases}
2.6\, v_{\text{esc}} \left( \dfrac{Z_{\text{surf}}}{Z_{\odot}} \right)^{0.13}, & T_{\text{eff}} \geq T_{\text{jump}}, \\[8pt]
1.3\, v_{\text{esc}} \left( \dfrac{Z_{\text{surf}}}{Z_{\odot}} \right)^{0.13}, & T_{\text{eff}} < T_{\text{jump}},
\end{cases}
\end{equation}
where $Z_{\odot}$ is the solar metallicity, $v_{\text{esc}}=\sqrt{2 G M (1 - \Gamma_e)/R_*}$, with $M$, $R_*$, and $\Gamma_e$ denoting the stellar mass, photospheric radius, and electron-scattering Eddington factor, respectively.
The temperature $T_{\text{jump}}$ corresponds to the bistability jump \cite{2001A&A...369..574V}.
\item WR stars are characterized by dense, optically thick winds, C/N/O-enriched surfaces, and strong wind clumping.
Since there is currently no universally accepted theoretical framework for WR wind acceleration, we adopt the empirical scaling proposed by \citet{nugisMasslossRatesWolfRayet2000}, which relates $v_\infty$ to the core escape velocity $v_{\text{esc,core}}$ derived from the hydrostatic core radius (see Section 6 in \citet{nugisMasslossRatesWolfRayet2000} for more details).
\end{itemize}

For collective winds, we assume that the outflows from different stars are fully mixed and attain a common velocity, with total momentum conserved during mixing and directional components of the momentum vectors neglected. Accordingly, the velocity of the collective winds can be expressed as,

\begin{equation}
V(t) = \frac{\int_{8\,M_{\odot}}^{M_{\mathrm{up}}} \dot{m}(M_0, t) \, v_\infty(M_0, t) \, \xi(M_0) \, dM_0}{\int_{8\,M_{\odot}}^{M_{\mathrm{up}}} \dot{m}(M_0, t) \, \xi(M_0) \, dM_0},
\end{equation}

where \(\dot{m}(M_0, t)\) denotes the mass-loss rate at time $t$ for a star with an initial mass $M_0$, $M_{\mathrm{up}}$ represents the maximum initial stellar mass, typically taken to lie in the range of $100$--$150\,M_{\odot}$ and the minimum initial mass is set to $8\,M_{\odot}$. Unless otherwise specified, we adopt $M_{\mathrm{up}}=150~M_{\odot}$ throughout this work. The function \(\xi(M_0)\) is the initial mass function (IMF) of MSC. Here we adopt the segmented power-law IMF proposed by \citet{2001MNRAS.322..231K}, 

\begin{equation}
\label{eq:IMF}
\xi(M_0)\propto M_0^{-2.3}~~ \text{for}~M_0\ge 0.5M_{\odot} 
\end{equation}


\begin{figure*}
    \centering
    \includegraphics[width=0.75\linewidth]{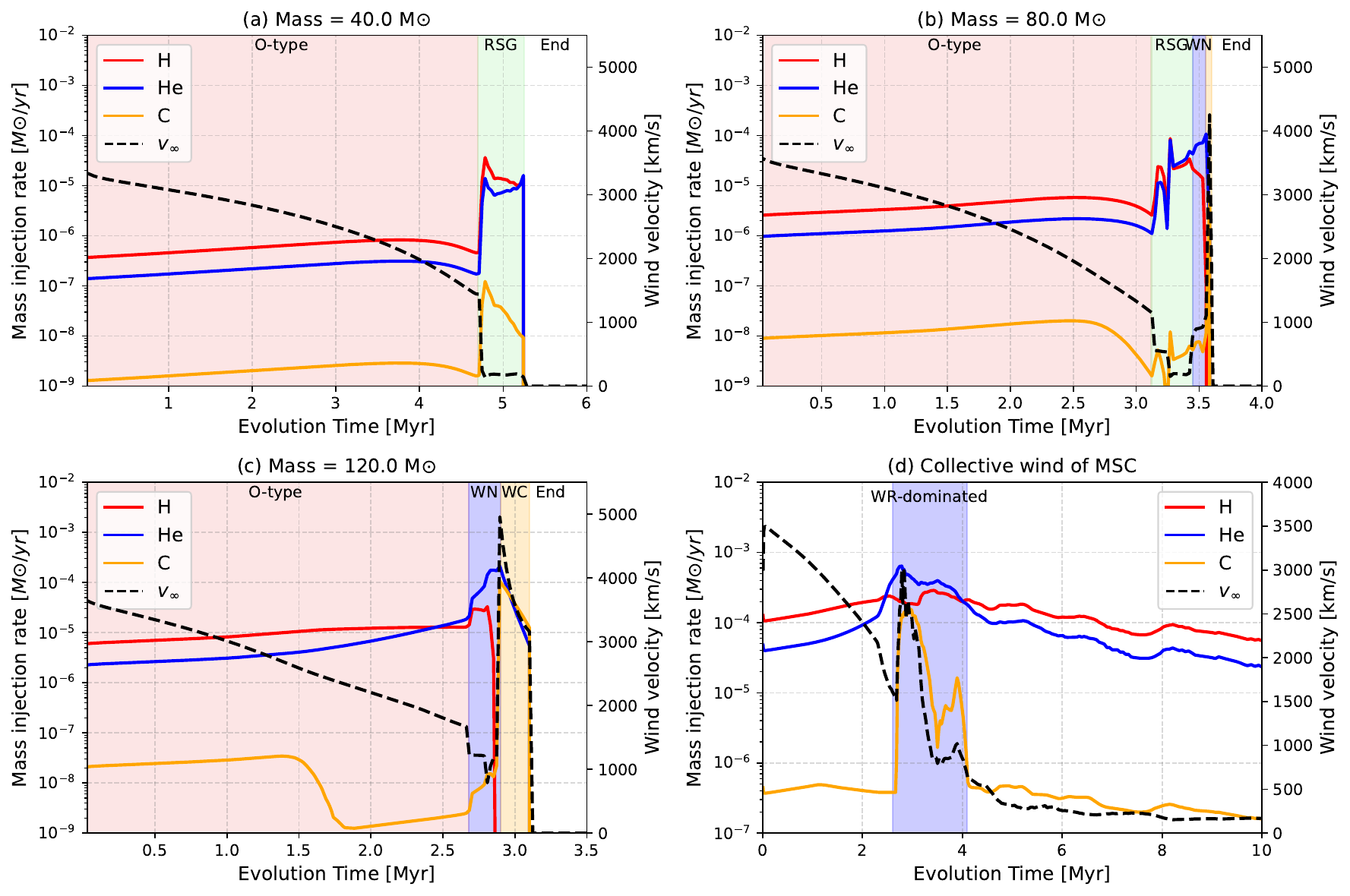}
    \caption{Temporal evolution of stellar wind velocity (dashed lines, right y-axis) and mass injection rates of H, He, and C (solid lines, left y-axis). Panels (a)–(c) show stars of 40, 80, and 120 $\,M_{\odot}$, respectively, panel (d) shows the collective winds. Colored regions indicate evolutionary stages, red for O-type, green for RSG, blue for WN-type, and orange for WC-type. Blank areas mark the end of stellar evolution.}
    \label{fig:StarEvo}
\end{figure*}

\begin{figure*}
    \centering
    \includegraphics[width=0.77\linewidth]{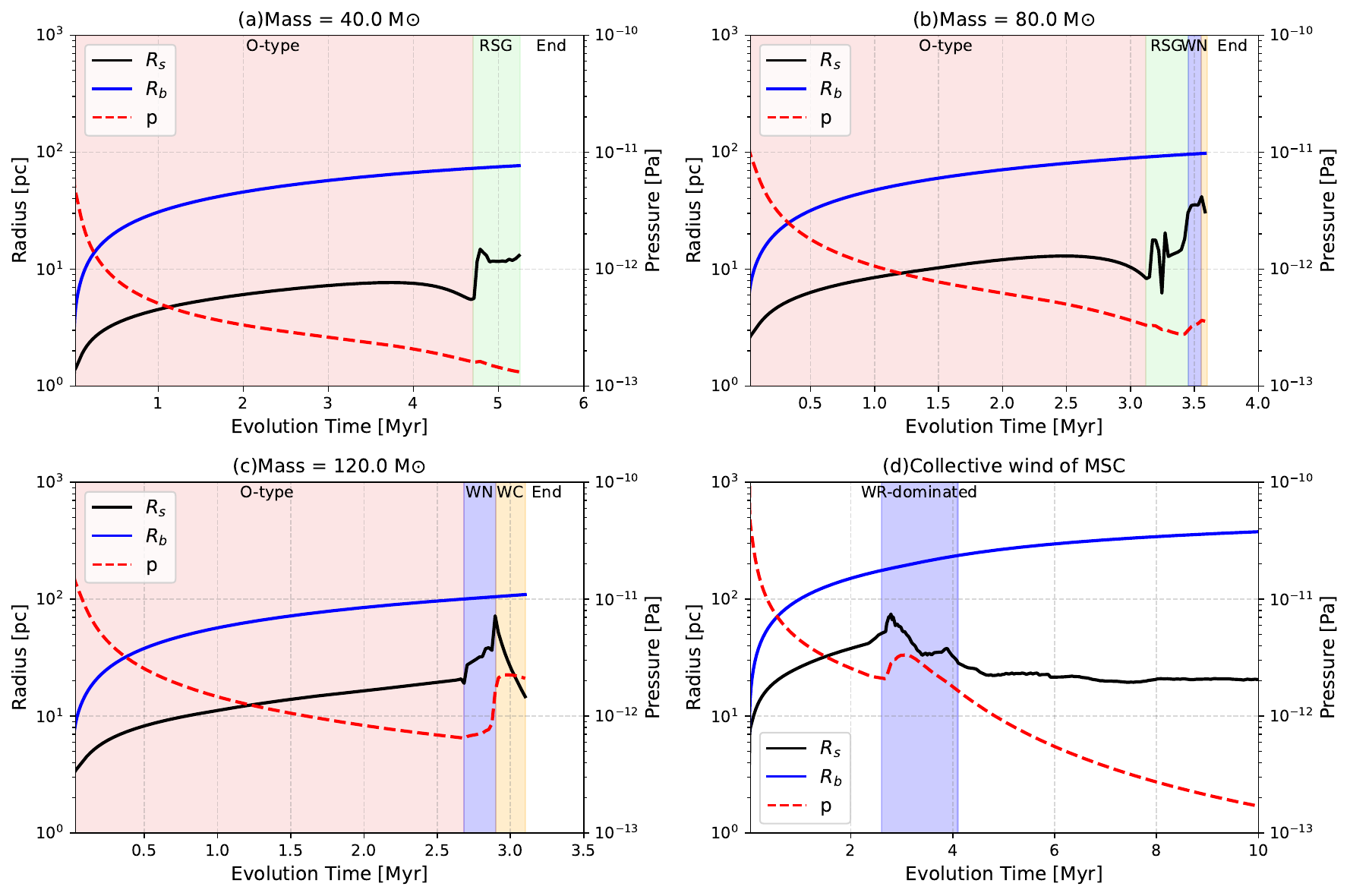}
    \caption{Temporal variations of $R_s$, $R_b$ and $p$ at the WTS. Solid and dashed lines correspond to the left and right y-axes, respectively. Colored regions and panel definitions are the same as in Fig.~\ref{fig:StarEvo}.}
    \label{fig:shock}
\end{figure*}

Figure~\ref{fig:StarEvo} presents the evolution of stellar winds from stars with initial masses of $40$, $80$, and $120\,M_{\odot}$, including their terminal wind velocities and the mass-loss rates of various elements, obtained with MIST, and Fig.~\ref{fig:StarEvo}d shows that of the collective winds of MSC. Shaded regions with different colors highlight the various evolutionary phases. The main stellar-wind phases and their characteristic properties are summarized below.
\begin{itemize} 
\item Following their formation, massive stars initially evolve along the main sequence as O-type stars. During this phase, the gradual expansion of the stellar radius leads to a decrease in \(v_\infty\), while the stellar wind remains dominated by hydrogen, as shown in the red shaded region in Fig.~\ref{fig:StarEvo}.

\item In the late main-sequence evolutionary stage, stars evolve into the red supergiant (RSG) phase. Other transitional stages, such as blue supergiants and luminous blue variables, exhibit similar wind properties and are not distinguished here. During this phase, the stellar envelope undergoes rapid expansion, accompanied by intense mass loss. As a consequence, the stellar wind becomes powerful but possesses an extremely low \(v_\infty\), as seen in green shaded region in Fig.~\ref{fig:StarEvo}.

\item Once the hydrogen-rich envelope is stripped away, the stars enter the WR phase. At this stage, the stellar winds are extremely strong and characterized by high outflow velocities and an almost complete absence of hydrogen.
The WR phase can be further divided into two subtypes, Wolf–Rayet Nitrogen (WN) stars, whose winds are dominated by helium, and Wolf–Rayet Carbon (WC) stars, whose winds are primarily composed of carbon and oxygen (blue and orange shaded regions in Fig.~\ref{fig:StarEvo}). The wind strengths of these two subtypes differ significantly, with WC-type stars typically having higher mass-loss rates and terminal velocities.
\end{itemize}

The stellar lifetime decrease with increasing initial mass. Most massive stars with $M_0 \lesssim 60M_{\odot}$ terminate their evolution in the RSG phase (Fig.~\ref{fig:StarEvo}a), during which the stellar winds remain hydrogen dominated throughout the entire evolution. By contrast, very massive stars with $M_0 \gtrsim 60M_{\odot}$ evolve successively through the O-star, RSG, and WR phases, despite their short lifetimes ($<$ 4 Myr). During the WR phase, stars further evolve through the the WN and WC subtypes, with the relative duration of the WC subtype increasing with initial mass. For instance, for an 80 $M_{\odot}$ star (Fig.\ref{fig:StarEvo}b), the WR phase is dominated by the WN subtype, while the WC subtype appears only briefly near the end of the evolution. By contrast, for a 120 $M_{\odot}$ star (Fig.\ref{fig:StarEvo}c), the WN and WC phases have nearly comparable durations.

For collective winds, stars within a cluster can simultaneously occupy different evolutionary stages, which smooths out the sharp transitions characteristic of single-star evolution. Nevertheless, because of the short lifetimes of massive stars, members of a coeval cluster begin to enter the WR phase collectively at an age of $\sim$\(2.6\,\mathrm{Myr}\). This transition systematically modifies the collective wind properties, leading to enhanced mass loss, higher wind velocities, and substantial enrichment of helium, carbon, and oxygen in the outflow (blue shaded region in Fig.~\ref{fig:StarEvo}d).

\subsection{Dynamical Evolution of Wind Termination Shocks
}\label{subsec:evolution_of_WTS}
The maximum energy attainable by CRs is estimated by modeling the time-dependent expansion of the WTS.
We adopt the hydrodynamic framework of \citet{weaverInterstellarBubblesII1977}, while neglecting radiative losses, to describe the coupled evolution of the shock radius \(R_s\) and the radius of the wind-driven bubble \(R_b\) (as shown in Fig.~\ref{fig:Schematic}). In simple terms, $R_s$ is determined by the balance between the ram pressure of the stellar wind and the internal pressure of the wind-driven bubble, while $R_b$, due to the large amount of ISM swept up by the forward shock, is governed by the following dynamical evolution equation:

\begin{equation}
\begin{aligned}
&R_s = \left(\frac{F(t)}{4\pi p}\right)^{1/2}, \\
&E_b = 2\pi p\,(R_b^3 - R_s^3), \\
&\frac{d}{dt}\left(\frac{4\pi \rho_0}{3}R_b^3\frac{dR_b}{dt}\right) = 4\pi R_b^2\,(p - p_{\mathrm{II}}), \\
&\frac{dE_b}{dt}= L(t) - 4\pi R_b^2 p \frac{dR_b}{dt}.
\end{aligned}
\label{eq:WTS_evolution}
\end{equation}

Here \(E_b\) is the total energy contained in the bubble, \(p\) the internal pressure, \(\rho_0\) the density of the ambient ISM, and \(p_{\mathrm{II}}\) the pressure of the surrounding H II region ionized by massive star radiation. In this work, we take the typical ISM number density $n_0 =1\ cm^{-3}$ and the temperature of the H II region $T = 8000~K$. The quantities $F(t)$ and $L(t)$ represent the momentum flux and mechanical luminosity of the stellar winds, respectively.
For a WTS driven by a single star with initial mass of \(M_0\), they are expressed as \(F(M_0,t) = \dot{m}(M_0,t) v_\infty(M_0,t)\) and \(L(M_0,t) = \frac{1}{2} \dot{m}(M_0,t) v_\infty^2(M_0,t)\).
For a collective wind, the total momentum and kinetic-energy input rates are obtained by integrating over \(\xi(M_0)\):
\begin{equation}
\begin{aligned}
F(t) &= \int_{8M_\odot}^{M_{\mathrm{up}}} \dot{m}(M_0,t) \, v_\infty(M_0,t) \, \xi(M_0) \, dM_0, \\
L(t) &= \int_{8M_\odot}^{M_{\mathrm{up}}} \frac{1}{2} \, \dot{m}(M_0,t) \, v_\infty^2(M_0,t) \, \xi(M_0) \, dM_0.
\end{aligned}
\label{eq:FL}
\end{equation}

Since $v_{\infty}(M_0,t)$ and $\dot{m}(M_0,t)$ vary significantly across different evolutionary phases, Eq.
\ref{eq:WTS_evolution} does not admit a steady‑state analytical solution.
We therefore solve the system numerically.
The integration is initialized at \(t_0 = 10^{4}\,\text{yr}\) using the early-stage solutions of \citet{weaverInterstellarBubblesII1977}. The resulting temporal evolution is shown in Fig.~\ref{fig:shock}. 

For WTS driven by the stellar wind, during the O-type phase, both \(v_\infty(M_0,t)\) and \(\dot{m}(M_0,t)\) remain nearly constant. In this stage, the wind-driven bubble continues to expand because of its overpressured interior, leading to a gradual decrease in the internal pressure. As a result, the pressure balance at the wind termination shock is continuously readjusted, causing the termination shock radius to expand outward as well. Our calculations yield temporal scalings of \(R_s \propto t^{0.44}\) and \(R_b \propto t^{0.58}\), in good agreement with the theoretical expectations of~\citet{weaverInterstellarBubblesII1977}. After the star enters the RSG phases, the sharp increase in \(\dot{m}\) enhances the ram pressure of the stellar wind, driving the termination shock rapidly outward and increasing $R_s$ (green shaded region in Fig. \ref{fig:shock}). However, this phase contributes relatively little to the total energy $E_b$ and pressure $p$ of the wind-driven bubble, so the expansion rate of the outer bubble remains nearly unchanged. When the star subsequently evolves into the WR phase, both $\dot{m}$ and $v_{\infty}$ increase significantly, again causing rapid expansion of the termination shock. In contrast to the RSG stage, the WR wind injects a large amount of kinetic energy into the bubble, leading to significant increase in both $E_b$ and $p$. As a result, the internal pressure no long declines with time, which eventually drives the contraction of the termination shock near the end of the WR phase (orange shaded region in Fig.~\ref{fig:shock}c).

For the WTS driven by the collective winds of an MSC, the evolution is relatively simpler. Before the most massive stars enter the WR phase, the behavior of the termination shock and the wind-driven bubble resembles that of single stars in the O-type phase. Once the most massive members of the cluster reach the WR phase, the rising ram pressure of the collective wind initially dominates, driving rapid expansion of the termination shock. Subsequently, the increasing internal pressure of the bubble takes over as the main driver of the shock evolution. This sequence ends when the most massive stars complete their WR evolution. Thereafter, the bubble continues to expand, while its internal pressure gradually decreases. However, as the collective stellar wind weakens over time, the termination shock radius stabilizes, maintaining approximate equilibrium (see Fig.~\ref{fig:shock}d).

\subsection{Time-Dependent Cosmic-Ray Injection}
\label{subsec:injection_spectrum}

Based on the stellar evolution and dynamical evolution of WTS described above, we assume the time-dependent CR injection spectrum of a star with initial mass $M_0$, following a power-law distribution in rigidity, with an exponential cutoff at the maximum rigidity $R_{\text{max}}(t)$,

\begin{equation}
Q_{\mathrm{M_0}}(t,R,Z) = N_\mathrm{{Z,M_0}}(t) \left( \frac{R}{\mathrm{GV}} \right)^{\gamma} \exp\left(-\frac{R}{R_{\mathrm{max,M_0}}(t)}\right)
\label{eq:injection_spectrum}
\end{equation}

where $\gamma$ is the spectral index, assumed to be time-independent and identical for all particle species. Here $N_\mathrm{Z,M_0}(t)$ denotes the time-dependent injection rate of nuclei with charge \(Z\).
In this work, we consider only the dominant isotope of each element, and the subscript Z is used to label particle species unless isotopic distinctions are necessary. 

For collective winds, the contributions from stars of different initial masses have already been incorporated into $N_\mathrm{Z,M_0}$ and $R_{\mathrm{max,M_0}}$. The time-dependent average injection spectrum per massive star is then obtained by integrating over the IMF:
\begin{equation}
Q(t, R, Z) = \frac{\int_{8M_{\odot}}^{M_{\mathrm{up}}}\, Q_{M_0}(t,R,Z)\, \xi(M_0) \, dM_0}{\int_{8M_{\odot}}^{M_{\mathrm{up}}} \; \xi(M_0) \, dM_0}
\end{equation}

\subsubsection{Maximum Rigidity}
\begin{figure*}
    \centering
    \includegraphics[width=0.77\linewidth]{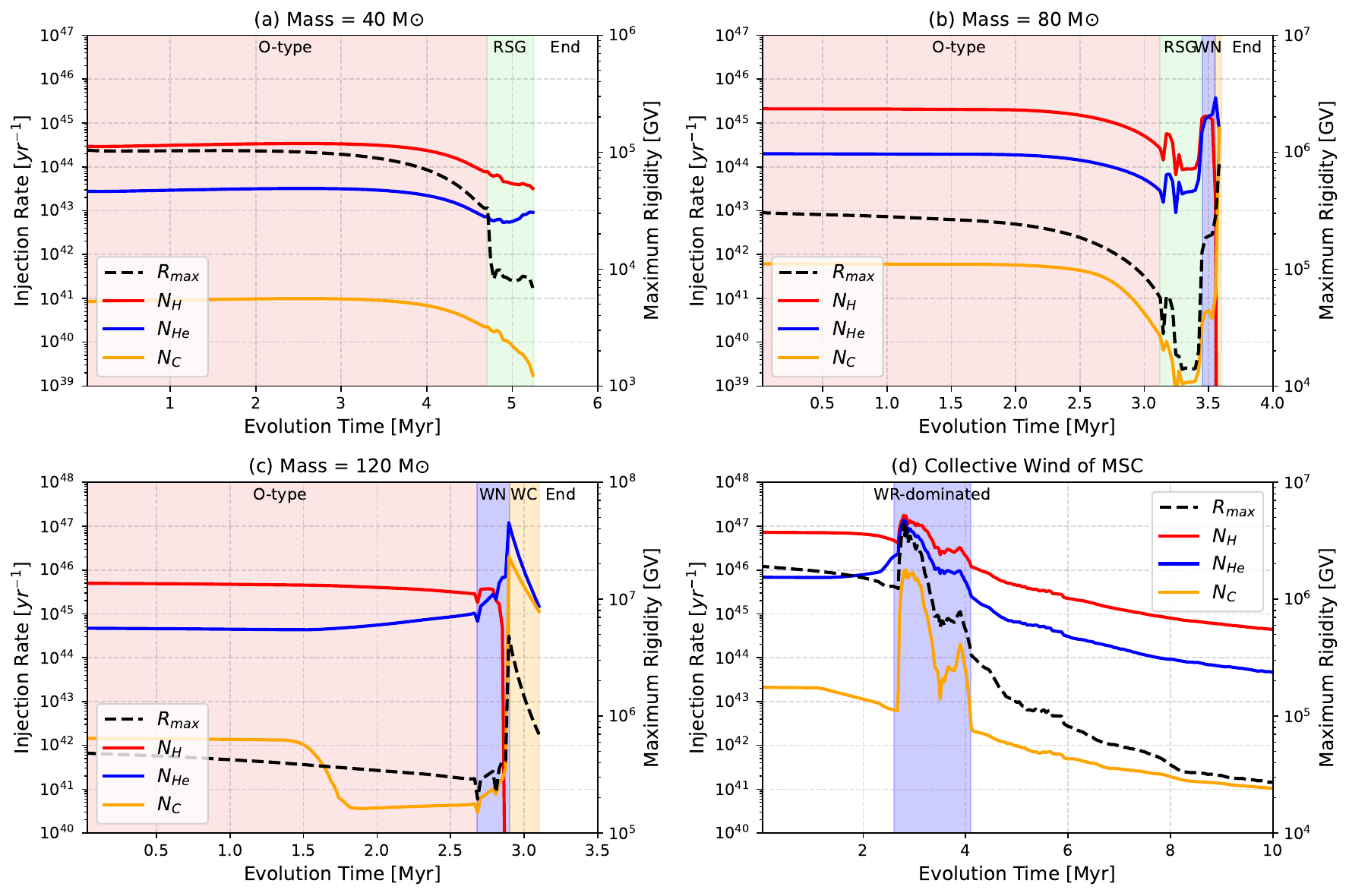}
    \caption{Temporal variations of $R_{\text{max}}$ and $N_Z$ with Bohm turbulence spectra ($\eta_B=0.01$, $\gamma=-2$, $\eta_{acc}=0.1$). Solid and dashed lines correspond to the left and right y-axes, respectively. Colored regions and panel definitions are the same as in Fig.~\ref{fig:StarEvo}.}
    \label{fig:injection_par}
\end{figure*}

In DSA, a particle's maximum achievable rigidity is constrained by its survival time within the shock, the system age \(t_{\rm age}\) or the escape timescale \(t_{\rm esc}\), since acceleration to a given energy requires a timescale $t_{\text{acc}}$, such as $t_{\text{acc}} \lesssim \text{min}(t_{\text{esc}},t_{\text{age}})$. 

For a shock with flow speed $V$ and particle diffusion coefficient $K$, the competition between diffusive escape from the shock and advection transport toward it produces a characteristic diffusion length scale of $K/V$.
The typical residence time of a particle on either side of the shock is given by the time required to diffuse across this scale, $t_{\text{res}} \sim (K/V)/c = K/(cV)$.
For a strong shock, the velocity jump satisfies $\Delta V \sim V$, so that the fractional energy gain per acceleration cycle (i.e., two shock crossings) is $\Delta E/E \sim V/c$.
Therefore, a particle requires approximately $c/V$ cycles to double its energy.  
The acceleration timescale is then obtained by multiplying the residence time per cycle by the number of cycles required for substantial energy gain, 
$t_{\text{acc}} \sim \left( \frac{K}{cV} \right) \times \left( \frac{c}{V} \right) = \frac{K}{V^2}.$

In the WTS system, the characteristic age $t_{\text{age}}$ is comparable to the lifetime of massive stars, i.e., of order several Myr. Assuming diffusive escape, the particle escape timescale can be estimated as $t_{\text{esc}} \sim R_s^2/K$. Adopting $K=1\times 10^{28}\ cm^2/s$ and $R_s =10\ pc$, we obtain $t_{\text{esc}}\approx 3\ kyr\ll t_{\text{age}}$. This indicates that particle escape, rather than the finite system age, provides the primary constraint on the maximum attainable rigidity. Using the rigidity dependence of the diffusion coefficient, \(K(R)\), we then obtain the relation \citep{malkovNonlinearTheoryDiffusive2001},
\begin{equation}
\frac{K(R_{\text{max}})}{V} \sim R_s.
\label{eq:rmax_condition}
\end{equation}

Here, \(K(R)\) is commonly parameterized as~\cite{morlinoParticleAccelerationWinds2021},
\begin{equation}
K(R) = \frac{c}{3} r_L \left( \frac{r_L}{L_c} \right)^{\delta},
\label{eq:diff}
\end{equation}
where \(r_L\equiv R/B\) is the Larmor radius, with $R$ denoting the particle rigidity and $B$ the magnetic field strength. The parameter \(L_c\) represents the turbulence coherence length, while the index $\delta$ depends on the assumed turbulence model.
Specifically, $\delta=0$ corresponds to the Bohm diffusion limit, which represents the strongest scattering regime \cite{Bohm}, $\delta=-1/2$ for Kraichnan MHD turbulence and $\delta=-2/3$ for the classical Kolmogorov cascade \cite{Kraichnan:1965zz, Kolmogorov}. 

Assuming a fraction $\eta_B$ of the stellar wind kinetic energy converting into turbulent magnetic energy, there is $B(r) = \frac{1}{r} \sqrt{\frac{\mu_0 \eta_B L}{2\pi V}}$. Substituting $K(R)$ and $B(R_s)$ into Eq. (\ref{eq:rmax_condition}) and (\ref{eq:diff}), we obtain $R_{max}$ for different turbulence models,
\begin{equation}
R_{\text{max}}(t) = 
\begin{cases} 
3 \sqrt{\frac{\mu_0 \eta_B L(t)V(t)}{2\pi}} &(\text{Bohm}) \\
\frac{9 R_s(t)}{c L_c} \sqrt{\frac{\mu_0 \eta_B L(t)V(t)^3}{2\pi}} & \text{(Kraichnan)} \\
\frac{27 R_s(t)^2}{c^2 L_c^2} \sqrt{\frac{\mu_0 \eta_B L(t)V(t)^5}{2\pi}} & (\text{Kolmogorov})
\end{cases}
\label{eq:Rmax}
\end{equation}

For the Bohm model, $R_{\text{max}}$ depends only on the terminal velocity $v_{\infty}$ and the kinetic-energy injection rate $L$. While under the Kraichnan and Kolmogorov turbulence spectra, $R_{\text{max}}$ becomes more sensitive to the wind properties and additionally depends on the radius of the WTS. As the WTS evolves dynamically, increases in the wind kinetic luminosity or terminal velocity generally lead to a larger $R_{\text{max}}$ (except during the RSG phase). Consequently, the variation amplitude of $R_{\text{max}}$ across different evolutionary stages follows the ordering {\rm Bohm} < {\rm Kraichnan} < {\rm Kolmogorov}.
Specifically, when a $M_0=120M_{\odot}$ star evolves from the O-type phase to the WC phase, $L$ increases by about two orders of magnitude, while both $v_{\infty}$ and $R_s$ increase by factors of $\sim$3. As a result, $R_{\text{max}}$ increases by about one order of magnitude in the Bohm case, but by about two and three orders of magnitude under the Kraichnan and Kolmogorov turbulence spectra, respectively.

For Bohm turbulence spectra, the time evolution of $R_{\text{max}}$ is shown in Fig.~\ref{fig:injection_par}. Under Kraichnan and Kolmogorov spectra, the evolution follows the same trends, differing only in the amplitude of variation. 
During the main-sequence phase, $R_{\text{max}}$ remains nearly constant. When the star enters the RSG phase, the substantial decrease in $v_{\infty}$ reduces $R_{\text{max}}$. During the subsequent WR phase, the sharp increase in wind kinetic power leads to a dramatic rise in $R_{\text{max}}$. 

By contrast, for WTS formed by collective wind, the evolutionary transitions of individual stars have little impact on the overall system (as shown in Fig.~\ref{fig:injection_par}d). A significant increase in $R_{\text{max}}$ occurs when the most massive stars collectively enter the WR phase, because of the rapid enhancement of the wind power during this stage. After these massive stars reach the end of their lifetimes (approximately 4 Myr after formation), the collective-wind power declines substantially, causing $R_{\text{max}}$ to decrease rapidly toward the end of the MSC evolution. The difference in $R_{\text{max}}$ between the early and late phases of the MSC can exceed one order of magnitude.

\begin{figure*}
    \centering
    \includegraphics[width=0.9\linewidth]{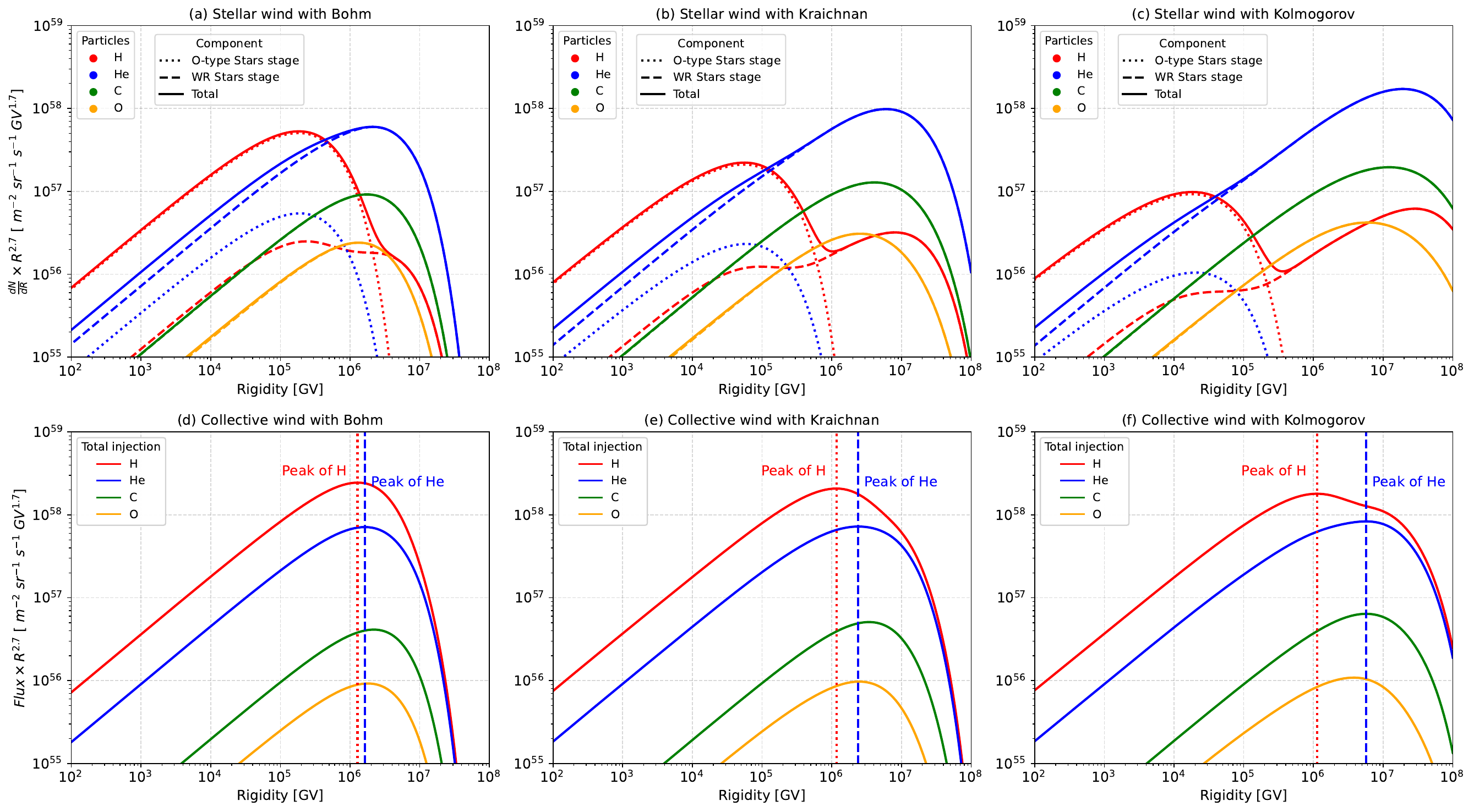}
    \caption{Total injection spectra of different particle species for a cluster with a total stellar mass of \(3\times10^4\,M_\odot\). Panels (a)–(c) and (d)–(f) show the stellar-wind and collective-wind models, respectively. From left to right, the columns correspond to the Bohm, Kraichnan, and Kolmogorov turbulence models. Colors indicate particle species, and dotted, dashed, and solid lines denote the contributions from the O-type phase, WR phase, and the total spectra, respectively.}
    \label{fig:inj_stellarwind}
\end{figure*}

\subsubsection{Injection Rate}
In DSA, a fraction of the kinetic energy of the stellar wind is transferred to high-energy particles (R > 1 GV), defined as $\eta_{acc}$. Throughout this work, $\eta_{acc}$ is assumed to be constant and identical for all particle species. The normalization factor for any massive star ($N_Z(t)$) in Eq. \ref{eq:injection_spectrum} is the ratio between the total injected energy and the average energy per accelerated particle, 
\begin{equation}
N_\mathrm{Z}(t) = \frac{\eta_{acc}\ n_Z(t) L(t)}{\langle E(t)\rangle},
\end{equation}

where $n_\mathrm{Z}(t)$ is the relative number density of the element $Z$, obtained from MIST, and $\langle E(t)\rangle$ is the average energy of high-energy particles,
\begin{equation}
\langle E(t)\rangle = \frac{\int_{1GV}^{\infty} \left( \sum_Z n_\mathrm{Z}(t) E_k \right) R^{\gamma} e^{-R/R_\mathrm{max}(t)} dR}{\int_{1GV}^{\infty} R^{\gamma} e^{-R/R_\mathrm{max}(t)} dR},
\end{equation}

where $E_k= \sqrt{(Z e R)^2 + m_Z^2} - m_Z$ is the kinetic energy of the element $Z$ with rigidity $R$.

Figure~\ref{fig:injection_par} illustrates the temporal evolution of $N_Z$ for 40, 80 and 120 $M_\odot$, whose trend is largely consistent with the mass-loss rates of different elements displayed in Fig.~\ref{fig:StarEvo} (except during the RSG phase). Once stars enter the WR stage, the injection rate increases rapidly, accompanied by a significant decline in the hydrogen abundance. Subsequently, helium becomes the dominant component. In the case of supermassive stars, the carbon abundance rises markedly after the onset of the WC stage. For the collective wind, after the most massive stars in the cluster enter the WR phase, their winds dominate the collective outflow. At this stage, the wind becomes strongly enriched in He, C, and O, contributing the majority of these elements released throughout the lifetime of the MSC.

\subsection{Spectral Features of the Total Injection}

We perform a time integration to obtain the total injection spectrum for an MSC of total mass $3\times 10^4\,M_{\odot}$, as shown in Fig.~\ref{fig:inj_stellarwind}. All calculations use the same parameters, $\gamma=-2$, $\eta_{acc}=0.1$, and $\eta_B=0.01$.

As expected, the integrated stellar-wind injection spectrum naturally separates into two components from O-type and WR stars (dotted and dashed lines in Fig.~\ref{fig:inj_stellarwind}). The O-star component is proton-dominated and exhibits a lower cutoff rigidity, whereas the WR-star component is enriched in He, C, and O and can accelerate particles to much higher energies. This bimodal structure indicates that within the stellar-wind acceleration framework, a universal rigidity injection spectrum for all species is no longer valid, highlighting a key difference from previous work. In the collective-wind scenario, mixing among individual stellar winds suppresses this bimodal feature, leaving only a modest difference between H and He, particularly under Bohm diffusion. 

If such an intrinsic bimodality were present in the observed spectra, it would be in tension with the LHAASO results. We therefore examine whether the wind scenarios considered here can alleviate this inconsistency. Among the different turbulence prescriptions, the H-He difference is smallest under the Bohm diffusion in both stellar-wind and collective-wind environments. If the predicted spectra remain inconsistent with observations even in the Bohm limit, the corresponding scenario is unlikely to be the dominant PeVatron candidate. We therefore restrict the following analysis to the Bohm diffusion case.



\section{Tests of Stellar and Collective Wind PeVatron Models}
\label{sec:result}
We assume that particle diffusion in the Milky Way depends only on rigidity, independent of particle species, corresponding to the standard diffusion model. As a result, differences in the source rigidity spectra among particle species are preserved, and their characteristics are directly reflected in the observed CR spectra. We use the latest LHAASO observations to examine whether the two stellar-wind scenarios can act as dominant PeVatrons and account for CRs around the knee.

\subsection{Spectral Break at the Knee Region}
The first criterion examines whether the injection spectra of H and He share the same rigidity cutoff, corresponding to the knee feature. Owing to the time dependence of the injection spectrum, the integrated spectrum deviates from the standard power-law distribution with an exponential cutoff, whose local spectral slope is
\begin{equation}
    \frac{d(lnQ)}{d(lnR)}= \gamma-\frac{R}{R_{max}}
\end{equation}

\begin{figure}[htbp]
    \centering
    \includegraphics[width=0.76\linewidth]{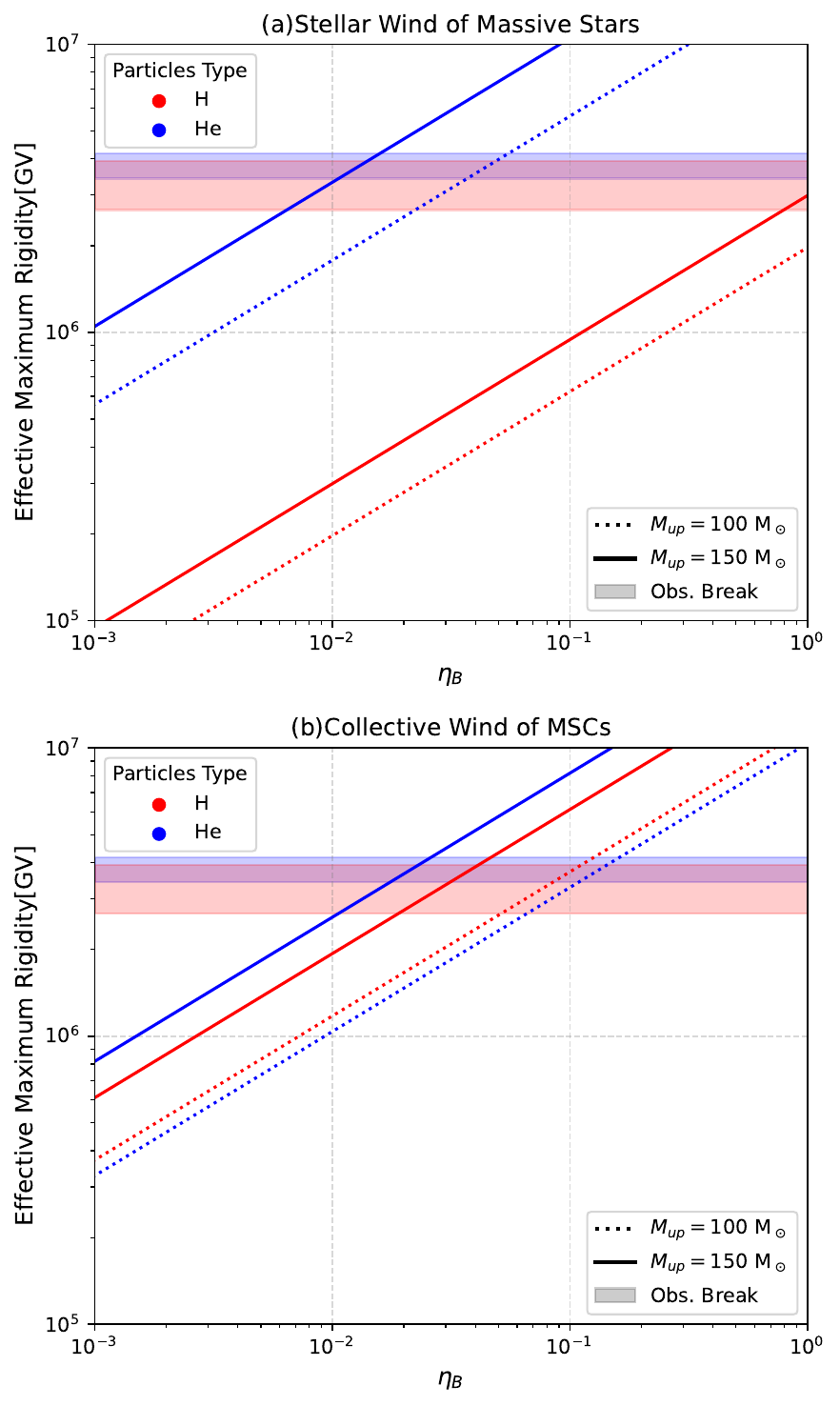}
    \caption{Effective maximum rigidity of H and He for different values of $\eta_B$. Panels (a) and (b) show the results for particle injection from individual massive-star winds and collective winds in MSCs, respectively. Red and blue lines correspond to H and He. The shaded bands indicate the break rigidity and its uncertainty measured by LHAASO \cite{collaborationFirstIdentificationPrecise2025,collaborationPreciseMeasurementCosmic2025}. Solid and dashed lines represent the numerical results for $M_{\mathrm{up}}=150\,M_{\odot}$ and $100\,M_{\odot}$, respectively.}
    \label{fig:eff_max}
\end{figure}

We define the effective maximum rigidity $R_{\mathrm{max}}^{\mathrm{eff}}$ as the rigidity at which the local spectral slope steepens by one relative to the injection index, i.e., $d(lnQ)/d(lnR)=\gamma-1$. The value of $R_{\mathrm{max}}^{\mathrm{eff}}$ depends on both $\eta_B$ and $M_{\mathrm{up}}$. A larger $M_{\mathrm{up}}$ enhances the contribution from WR stars, while a higher $\eta_B$ corresponds to more efficient particle confinement in the shock. Both factors lead to a larger $R_{\mathrm{max}}^{\mathrm{eff}}$, as shown in Fig.~\ref{fig:eff_max}.

For stellar winds of massive stars, proton and helium nuclei exhibit a substantial difference in $R_{\mathrm{max}}^{\mathrm{eff}}$. Even adopting a low $M_{\mathrm{up}}$ to suppress the contribution from WR stars, the maximum rigidity of proton and helium remains inconsistent with observational constraints, as shown in Fig.~\ref{fig:eff_max}a. Another constraint is that $\eta_B$ required to accelerate CRs up to the knee should lie within a reasonable range, typically around 0.1. For injection from O-type stars, even with $M_{\mathrm{up}} = 150\,M_{\odot}$, $\eta_B$ needs to approach 1 to reach the spectral break energy. 

For collective winds, although a noticeable difference in $R_{\mathrm{max}}^{\mathrm{eff}}$ between proton and helium persists even for $M_{\mathrm{up}}=150M_{\odot}$, the discrepancy remains within observational uncertainty (Fig.~\ref{fig:eff_max}b). In addition, $\eta_B$ needed to accelerate CRs to knee energies remains within a reasonable range. Consequently, collective winds in MSCs remain a viable PeVatron scenario.

\subsection{Elemental Abundances at the Knee Region}

During the propagation of CRs in the Milky Way, their interactions with the ISM produce secondary particles, which significantly modify the observed H/He ratio at low energies. However, the contribution of secondary particles to the knee region is negligible. Consequently, if PeV CRs arise from a single source population, the observed H/He ratio should match the injected ratio at the source.

\begin{figure}[htbp]
    \centering
    \includegraphics[width=0.8\linewidth]{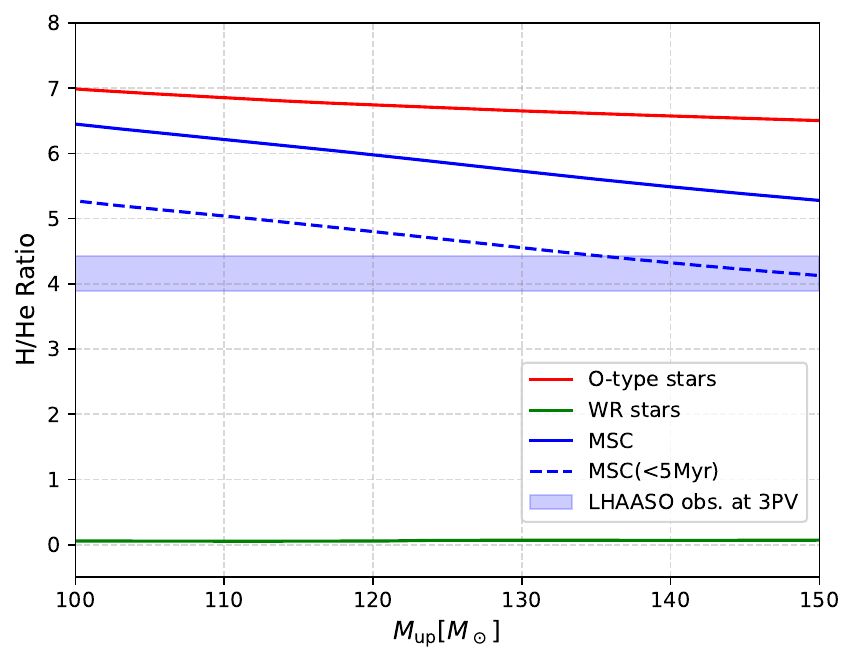}
    \caption{Comparison of the H/He ratio from different injection sources with observational results \cite{collaborationPreciseMeasurementCosmic2025}. The shaded region represents the LHAASO measurement at 3 PV and its uncertainty. Solid lines in different colors correspond to distinct injection sources, while the blue dashed line denotes the contribution from collective winds within 5 Myr.}
    \label{fig:H-He-ratio}
\end{figure}

We calculate the injected H/He ratio for stellar winds under different stellar mass upper limits, and compare the results with the LHAASO measurements, as shown in Fig.~\ref{fig:H-He-ratio}. The H/He ratio predicted for WR-star winds shows a clear discrepancy with the observations. For O-type stellar winds, the proton fraction is relatively higher, and the tension can be partially alleviated by adopting a harder rigidity injection spectrum for helium. For collective winds, the predicted H/He ratio also exceeds the observed value. However, as discussed in Section \ref{sec:methods}, collective winds can accelerate particles to higher rigidities during the early evolutionary phase (age $<$ 5 Myr). Restricting the injection to this phase yields an H/He ratio broadly consistent with the observations, without requiring any additional corrections (dashed blue line in Fig.~\ref{fig:H-He-ratio}).

We summarize all test outcomes in Table \ref{tab:test}. It is evident that stellar winds of massive stars, whether from WR stars, O-type stars, or a combination of both, are unlikely to serve as primary PeVatron sources. In contrast, the collective winds of MSCs exhibit good consistency with the observational results in all tests.

\begin{table}[htbp]
\centering
\caption{Test Results of Different Sources as the Primary PeVatron.}
\label{tab:test}
\begin{tabular}{|l|c|c|c|}
\hline
Source Type & Break of Knee& $\eta_B$ limitation & H/He\\
\hline
MSCs & $\checkmark$&$\checkmark$&$\checkmark$ \\
O-type + WR stars & $\times$&$-$&$-$ \\
O-type stars & $\checkmark$&$\times$&$\checkmark$ \\
WR stars & $\checkmark$&$\checkmark$&$\times$ \\
\hline
\end{tabular}
\begin{tablenotes}
    \footnotesize
    \item For the case of O‑type + WR stars, the maximum rigidities of proton and helium differ significantly, rendering the other two tests complicated and unnecessary.
\end{tablenotes}
\end{table}

\subsection{Total Cosmic-Ray Energy Budget}

Although the predictions for collective winds agree well with the observations, it remains unclear whether MSCs can provide sufficient energy to account for the measured CR spectrum. As the dominant accelerator for CRs below 100 TeV, SNRs typically release a total energy of $10^{51}\,\mathrm{erg}$ per explosion, with a Galactic explosion rate of approximately one event per century. This corresponds to a total energy injection rate of about $10^{49}\,\mathrm{erg/yr}$. For an MSC with a total mass of $3 \times 10^4\,M_{\odot}$, the total injected energy is approximately $10^{53}\,\mathrm{erg}$, obtained by integrating $L$ in Eq. (\ref{eq:FL}). The total mass $M_{cl}$ of star clusters follows an IMF scaling as $M_{cl}^{-2}$~\cite{2019ARA&A..57..227K}. Since we assume that only MSCs with $M_{cl}\ge10^4\,M_{\odot}$ produce collective winds, the total energy injection rate from MSCs can be estimated as,

\begin{equation}
    L_{\mathrm{MSC}}= \frac{M_{\mathrm{SFR}}\times 10^{53}\,\mathrm{erg}}{\ln(M_{\mathrm{max}}/M_{\mathrm{min}})}\left(\frac{1}{10^4\,M_{\odot}}-\frac{1}{M_{\mathrm{max}}}\right)
\end{equation}

where $M_{\mathrm{SFR}}$ is the average star formation rate of the Milky Way, $M_{\mathrm{min}}$ and $M_{\mathrm{max}}$ denote the lower and upper bounds of the star cluster mass. For $M_{\mathrm{SFR}}=2\,M_{\odot}\,\mathrm{yr}^{-1}$ \cite{2022ApJ...941..162E}, $M_{\mathrm{min}}=10^2\,M_{\odot}$\cite{2010ApJ...725.1886L}, and $M_{\mathrm{max}}=10^5\,M_{\odot}$, we obtain $L_{\mathrm{MSC}} \approx 4\times 10^{48}\,\mathrm{erg/yr}$ (40\% of $L_{\mathrm{SNR}}$). This comparable injection luminosity implies that MSCs possess a harder injection spectrum than SNRs, they can contribute sufficient PeV CRs to dominate the knee region.

We also note that star clusters with masses below $10^4\,M_{\odot}$ also release non-negligible energy via stellar winds, at a rate of approximately $10^{49}\,\mathrm{erg/yr}$. From the perspective of energy injection alone, stellar winds should contribute considerably to the observed CR spectrum. 

\section{A Stellar-dominated Model for Galactic Cosmic Rays} 
\label{sec:CWM}
Under the constraints imposed by LHAASO, only collective winds from MSCs can serve as dominant PeVatrons in the stellar-wind scenario. Although individual stellar winds are in principle also efficient CR accelerators, their maximum rigidity is substantially lower than that achieved by collective winds. This hierarchy in rigidity points to a multi-population origin of the CR spectrum.

We therefore propose a stellar-dominated model. Below 10 TeV, the spectrum remains dominated by SNRs. Individual massive stars can accelerate CRs up to several hundred TV, while MSCs, representing the collective counterpart of massive stars, extend the acceleration limit into the PV regime. Within this model, spectral transitions naturally emerge at intermediate energies, providing a unified explanation for the observed spectral hardening above 200 GeV and above 0.1 PeV. Furthermore, the distinct elemental abundances associated with different source populations may intrinsically explain the rigidity-dependent spectral differences observed among various particle species.

To assess the implications of this model and explore its characteristic properties, we perform numerical calculations using GALPROP v57~\cite{2022ApJS..262...30P}. Given the substantial uncertainties associated with the properties and evolution of massive stars and stellar clusters, parameter fitting would be highly degenerate and therefore of limited physical relevance. Instead, we present a representative set of numerical results together with the adopted GALPROP parameter configuration. We show that this model naturally reproduces the rigidity-dependent spectral differences among various particle species around 100 GV and 0.1 PV, without introducing additional species-dependent degrees of freedom.

\subsection{GALPROP Setup}
\label{subsec:galprop}

\begin{figure*}
    \centering
    \includegraphics[width=0.7\linewidth]{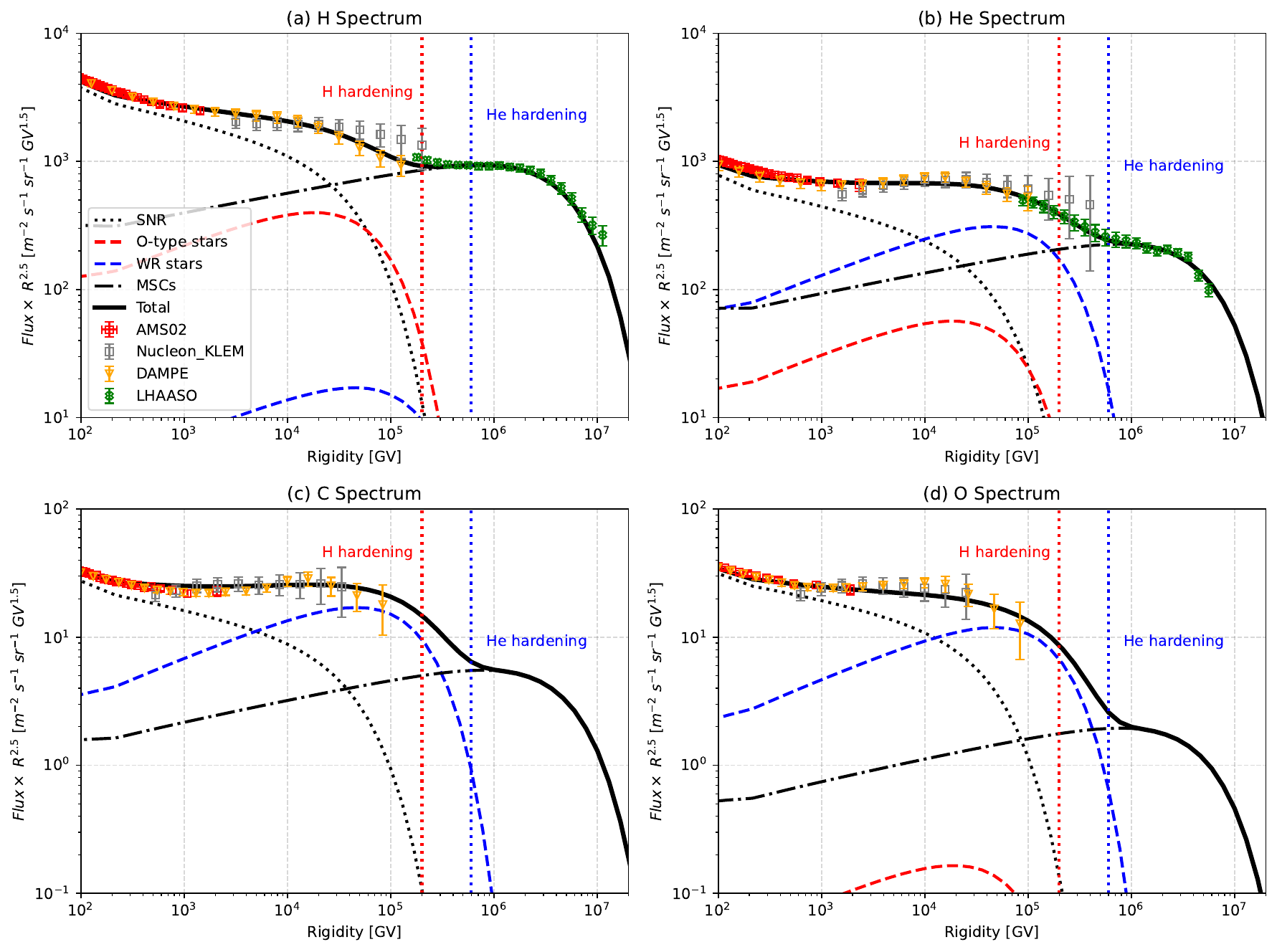}
    \caption{Comparison of particle rigidity spectra calculated with GALPROP and observations \cite{2021PhR...894....1A,2023PhRvL.130u1002A,2019AdSpR..64.2546G,2026NatAs.tmp...79L,collaborationFirstIdentificationPrecise2025,collaborationPreciseMeasurementCosmic2025}. Panels (a)--(d) show the spectra of protons, helium, carbon, and oxygen, respectively. Different line styles denote contributions from different source populations. The red and blue dashed lines represent the contributions from O-type and WR stellar winds, respectively. The approximate hardening rigidities of protons and helium are indicated by the red and blue dotted lines.}
    \label{fig:H-He}
\end{figure*}

The propagation of CRs is governed by the standard diffusion equation, which includes diffusion, convection, re-acceleration, energy losses, fragmentation, and radioactive decay. We adopt the propagation setup of Model B' from \citet{maInterpretationsCosmicRay2023}, which incorporates an additional high-energy break in the diffusion coefficient to reproduce the DAMPE-measured hardening of B/C and B/O ratios. The diffusion coefficient takes the form

\begin{equation}
D_{xx}(\mathcal{R}) =
\begin{cases}
D_0 \beta^\eta \left( \dfrac{R}{4\,\mathrm{GV}} \right)^\delta, & R \le R_h, \\[8pt]
D_0 \beta^\eta \left( \dfrac{R_h}{4\,\mathrm{GV}} \right)^\delta \left( \dfrac{R}{R_h} \right)^{\delta_h}, & R > R_h,
\end{cases}
\end{equation}

where $\beta$ is the particle velocity in units of $c$, $\delta$ and $\delta_h$ describe the interstellar turbulence properties below and above the break rigidity $R_h$, and $\eta$ is a phenomenological low-energy correction. Using this model, we neglect convection and directly adopt the best-fit parameters from \citet{maInterpretationsCosmicRay2023}. Solar modulation is treated via the force-field approximation \cite{gleesonSolarModulationGalactic1968}.

All source populations are assumed to follow the same spatial distribution as SNRs, as described in \citet{yuanPropagationCosmicRays2017}. Cosmic rays accelerated by stellar and collective winds are injected with a power-law spectrum with an exponential cutoff, given by Eq.~\eqref{eq:injection_spectrum}. For SNRs, we adopt a broken power law with an exponential cutoff,
\begin{equation}
q(R) =
\begin{cases}
q_0 R^{\gamma_1} e^{-R/R_{\mathrm{max}}}, & R \le R_{\mathrm{br}}, \\[6pt]
q_0 R_{\mathrm{br}}^{\gamma_1} \left( \dfrac{R}{R_{\mathrm{br}}} \right)^{\gamma_2} e^{-R/R_{\mathrm{max}}}, & R > R_{\mathrm{br}}.
\end{cases}
\end{equation}

To avoid extra complexity from the bimodal injection spectrum of stellar winds, we decompose the stellar winds of massive stars into two components: proton-rich winds from O-type stars and helium-dominated winds from WR stars, with the latter possessing a higher $R_{\mathrm{max}}$. The injection spectrum parameters and elemental abundances for all populations are listed in Tables~\ref{tab:parameters} and \ref{tab:abundances}. Except for the SNR, elemental abundances are obtained by numerical integration using MIST. For collective winds, the integration is performed over [0, 5]~Myr, corresponding to the evolutionary phase with the highest $R_{\mathrm{max}}$ (see Fig.~\ref{fig:injection_par}). For individual stellar winds, stars are classified as WR stars when $X_H < 0.1$ and as O-type stars otherwise. We adopt $M_{\mathrm{up}} = 105\,M_\odot$ for individual stellar winds, as this yields better agreement with the observed carbon flux than the $M_{\mathrm{up}} = 150\,M_\odot$ case; it is physically plausible that stars in lower-mass clusters follow a lower upper mass threshold \cite{2006MNRAS.365.1333W}.

\begin{table}[htbp]
\centering
\caption{Spectral parameters for different components.}
\label{tab:parameters}
\begin{tabular}{lcccc}
\hline
Parameters & SNR & O-type/WR stars & MSCs \\
\hline
    $\gamma_1$  & $-2.2$ & $-1.95$  & $-2.1$ \\
$\gamma_2$  & $-2.45$ & --- & --- \\
$R_{br}$[GV]  & $10$ & ---  & --- \\
$R_{max}$[TV] & $50$ & $60/150$ & $5000$ \\
\hline
\end{tabular}
\end{table}
\begin{table}[htbp]
\centering
\caption{Elemental abundances for different components.}
\label{tab:abundances}
\begin{tabular}{lcccc}
\hline
Element & SNRs & O-type stars & WR stars & MSCs \\
\hline  
H      & $8.00\times10^{-1}$ & $8.73\times10^{-1}$ & $4.73\times10^{-2}$ & $7.99\times10^{-1}$ \\
He     & $1.78\times10^{-1}$ & $1.26\times10^{-1}$ & $8.66\times10^{-1}$  & $1.94\times10^{-1}$ \\
C      & $6.69\times10^{-3}$ & $1.91\times10^{-4}$ & $4.89\times10^{-2}$  & $4.80\times10^{-3}$ \\
O      & $8.48\times10^{-3}$ & $3.85\times10^{-4}$ & $3.48\times10^{-2}$  & $1.71\times10^{-3}$ \\
Mg     & $1.97\times10^{-3}$ & $3.43\times10^{-5}$ & $1.14\times10^{-4}$  & $4.11\times10^{-5}$ \\
Si     & $1.69\times10^{-3}$ & $3.26\times10^{-5}$ & $1.06\times10^{-4}$  & $3.89\times10^{-5}$ \\
S      & $3.30\times10^{-4}$ & $1.37\times10^{-5}$ & $4.00\times10^{-5}$  & $1.61\times10^{-5}$ \\
Fe     & $1.74\times10^{-3}$ & $3.95\times10^{-5}$ & $1.34\times10^{-4}$  & $4.74\times10^{-5}$ \\
\hline
\end{tabular}
\begin{tablenotes}
  \small
  \item Note: We perform the elemental normalization at the same rigidity rather than at the same kinetic energy per nucleon.
\end{tablenotes}
\end{table}

\subsection{H–He Spectral Hardening at $\sim 0.1 PV$}
\label{subsec:H-He}

At sub-PeV energies, LHAASO found that the helium spectrum is softer than the proton spectrum and exhibits hardening at higher rigidity \cite{collaborationPreciseMeasurementCosmic2025}. Phenomenological studies often accounted for this difference by assuming that protons and helium originate from separate sources \cite{2025arXiv251106733Y}, but the underlying astrophysical mechanism remains unclear. We suggest that contributions from stellar winds of massive stars may provide a viable solution.

In this scenario, TeV CRs are dominated by stellar winds of massive stars, as shown in Fig.~\ref{fig:H-He}. Helium is primarily accelerated by WR stars, which have a higher maximum rigidity than O-type stars. This component dominates the observed CR flux at tens of TV and is eventually overtaken by contributions from MSCs around 0.5 PV, producing the observed hardening of helium. By contrast, WR stars contribute negligibly to the proton flux, which is mainly accelerated by O-type stars with a lower maximum rigidity. Consequently, the proton spectrum is dominated by MSCs around 0.2 PV, leading to spectral hardening at lower rigidity than that of helium. 

WR stellar winds, rich in carbon and oxygen, suggest that the spectra of carbon and oxygen at PeV energies are expected to exhibit similar spectral behavior as helium, including comparable spectral indices and rigidity-dependent hardening, shown in Fig.~\ref{fig:H-He}. This prediction could be tested with future LHAASO measurements.

\subsection{Primary Cosmic Ray Grouping above 100 GV}

\label{subsec:C}
\begin{figure*}
    \centering
    \includegraphics[width=1\linewidth]{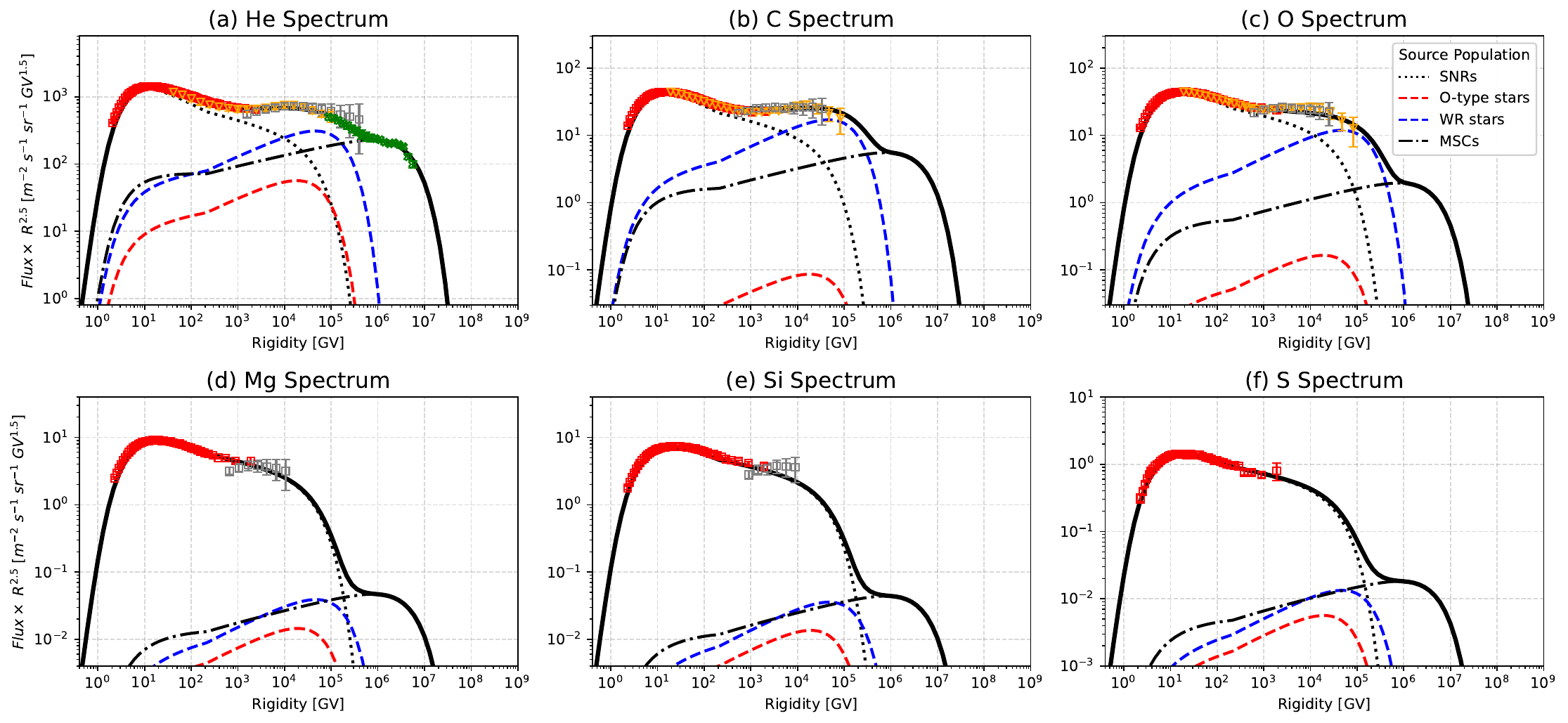}
    \caption{Comparison between calculated rigidity spectra and observations over a wide rigidity range \cite{2021PhR...894....1A,2023PhRvL.130u1002A,2019AdSpR..64.2546G,2026NatAs.tmp...79L,collaborationPreciseMeasurementCosmic2025}. Panels (a)--(c) in the first row show He, C, and O, which exhibit similar rigidity dependence, while panels (d)--(f) in the second row show Mg, Si, and S, which display different rigidity-dependent behavior. Line styles and colors follow the same notation as in Fig.~\ref{fig:H-He}.}
    \label{fig:group}
\end{figure*}

The Alpha Magnetic Spectrometer (AMS-02) has observed two distinct patterns of rigidity dependence for various elements above 100 GV \cite{2023PhRvL.130u1002A}. The first group includes He, C, O and Fe, while the second comprises Ne, Mg, Si and S. All these observed spectra are dominated by primary CRs, with only a small contribution from secondary particles such as those produced by spallation. Except for Fe, all elements have a charge-to-mass ratio of 0.5. Conventional models cannot account for the observed divergence in rigidity-dependent behaviors among these elements. In our scenario, differences in elemental abundances between SNRs and stellar winds provide a natural explanation for most elements，except for Fe. 

Above 100 GV, the observed CR fluxes of light nuclei (He, C, O) gradually shift from being dominated by SNR to being dominated by stellar winds, leading to progressive spectral hardening. By contrast, heavy nuclei (e.g., Ne, Mg, Si, S) are not substantially enriched in either O-type or WR stellar winds. As a result, their spectra remain largely unaffected by the stellar-wind component and do not exhibit the hardening observed in light nuclei, as shown in Fig.~\ref{fig:group}. Therefore, our model strongly predicts a significant difference between these two groups of particles in the TV energy range, which is expected to be verified by future DAMPE observations.

\begin{figure*}
    \centering
    \includegraphics[width=0.7\linewidth]{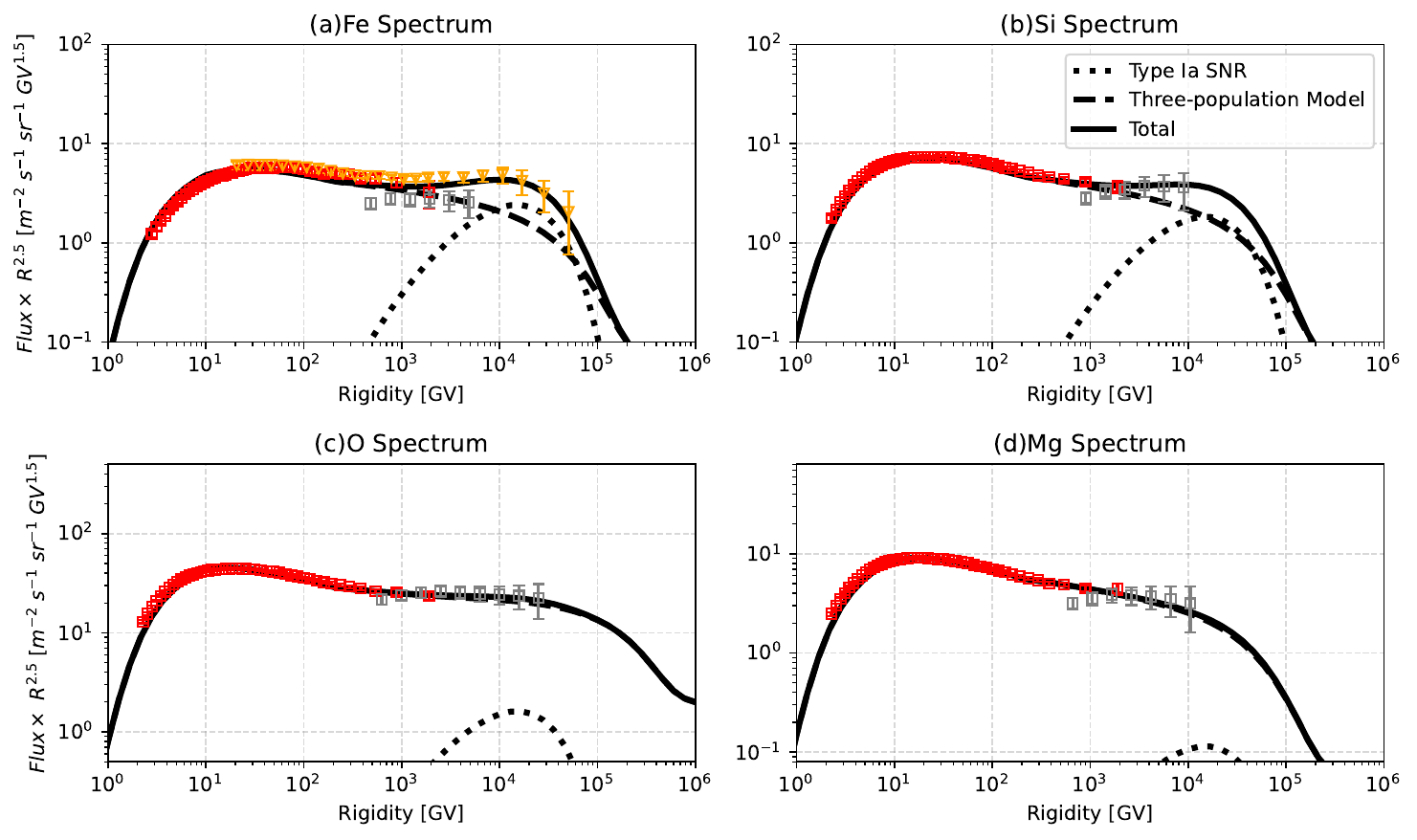}
    \caption{Rigidity spectra of various particles before and after adding the nearby Type Ia SNR component. Panels (a)--(d) show Fe, Si, O, and Mg, respectively. Dashed lines denote the theoretical results of the stellar-dominated model without the nearby source, dotted lines represent the contribution of the nearby Type Ia SNR, and solid lines correspond to the sum of both contributions.}
    \label{fig:Fe}
\end{figure*}

Similarly to heavy nuclei such as Mg and Si, Fe is also not significantly enriched in stellar winds. Consequently, our model predicts no obvious spectral hardening of Fe at $\sim 1$~TV, in tension with the latest DAMPE observations (see Fig.~\ref{fig:Fe}a). As the charge-to-mass ratio of Fe differs from that of other elements, introducing a modified Fe component (e.g., with a different spectral index) is physically motivated. An alternative scenario with a nearby Fe-rich source could be proposed, which should not significantly affect the energy spectra of other particles. Type Ia supernova, the primary source of Fe in the Milky Way, is the potential candidate. We adopt a three-dimensional spherical diffusion model to calculate the spectral contribution from a Type Ia supernova with an age of 50 kyr and at a distance of 1 kpc. The injection spectral index is set to $−2$, and the total energy of particles with rigidity above 1 GV is $5.5\times 10^{50}$ erg. The accelerated particles are assumed to originate primarily from material ejected during the supernova explosion, and the elemental abundances are adopted from the N100 model in \citet{2013MNRAS.429.1156S}.
 
As shown in Fig.~\ref{fig:Fe}, the supernova produces spectral hardening for Fe (Fig.~\ref{fig:Fe}a), Si (Fig.~\ref{fig:Fe}b), and S (similar to Si) at $\sim$ 1 TV, while leaving the spectra of other elements essentially unchanged (Fig.~\ref{fig:Fe}c and Fig.~\ref{fig:Fe}d). This additional source does not introduce conflicts with current observations of other elements. 

Although the two proposed scenarios do not encompass all possibilities, they are representative and do not produce spectral hardening of Mg at TV-scale rigidities. We therefore further expect that future precise measurements of the Mg spectrum by DAMPE can provide an effective test of the scenario's feasibility.

\section{Discussion and Conclusions}
\label{sec:dis}

In this work, we present the first time-dependent CR injection model that accounts for the full evolution of massive stars with the dynamical evolution of WTS. Our analysis shows that individual stellar winds produce different rigidity cutoffs for protons and helium, inconsistent with the spectral break near 3.5 PV observed by LHAASO. Consequently, acceleration by the winds of individual massive stars cannot explain the dominant PeVatron contribution at these energies. By contrast, the collective winds of MSCs naturally resolve this discrepancy by combining contributions from stars at different evolutionary stages, and thus remain viable PeVatron candidates.

We propose a stellar-dominated model for Galactic CRs in which SNRs dominate below 10 TV, individual stellar winds from O-type and WR stars provide the dominant contribution between 10 TV and a few hundred TV, and collective winds of MSCs dominate the knee region up to 4 PV. The model successfully reproduces the rigidity-dependent spectral features observed in different species around 100 GV and 0.1 PV, including the different hardening rigidities of proton and helium, without invoking species-dependent adjustments.

Our model yields two testable predictions for future LHAASO and DAMPE observations. First, the carbon and oxygen spectra should exhibit spectral hardening near 0.5 PV, similar to that observed in helium. Secondly, in the multi-TV transition region between SNRs and stellar winds, the Mg spectrum should not exhibit the same hardening as elements like He, C, and O. These predictions emerge naturally from the distinct enrichment patterns of WR winds and from the spectral transitions between different dominant source populations, offering a direct means of distinguishing the proposed model from alternative explanations.


Several simplifications in the present model require further investigation. Stellar rotation, initial metallicity, and binary evolution are known to significantly alter wind composition, mass-loss rates, and evolutionary timescales \cite{2013ApJ...764...21C,2020MNRAS.499..873S,2024ARA&A..62...21M}. Nevertheless, these uncertainties are unlikely to alter the fundamental difference between individual and collective winds. In addition, the $^{22}$Ne abundance is not provided in the MIST outputs, and the neon spectrum, a key diagnostic of WR stellar wind acceleration, is therefore not computed in this work. Future studies incorporating these physical processes and additional isotopes will help quantify the uncertainties in the injection spectra from stellar winds.

From an observational perspective, large MSCs such as Cyg OB2 \cite{collaborationUltrahighenergyGammarayBubble2024} and R136 \cite{HESS:2024xux} are well-established high-energy sources and potential contributors to Galactic diffuse emissions \cite{2026arXiv260321665E}. However, their complex internal environments make it difficult to unambiguously identify the dominant particle-acceleration mechanism. Low-mass star clusters, by contrast, are expected to be dominated by individual stellar winds and thus offer a cleaner window into wind acceleration, even though the low mechanical energy injection rate of individual winds makes their gamma-ray signature difficult to resolve. These systems therefore represent the most promising targets, and we suggest observing young, low-mass clusters ($\sim 10^3\,M_{\odot}$) free from SNR or pulsar contamination, such as RCW 32, 36, and 38 \cite{2024NatAs...8..530P}, to isolate the contribution of individual stellar winds to the TeV-PeV CR spectrum.

More broadly, if collective winds in MSCs are the dominant Galactic PeVatrons, their signatures should extend beyond CR spectra to high-energy gamma rays and neutrinos. Future multi-messenger observations, particularly by next-generation neutrino observatories operating in the TeV to PeV regime \cite{IceCube-Gen2:2021tmd, KM3NeT:2018wnd, TRIDENT:proposal, HUNT:2023mzt}, will provide a decisive test of this hypothesis and may ultimately establish MSCs as a major source of Galactic PeV particles.

\begin{acknowledgments}

We would like to thank Qiang Yuan, Yiqing Guo, Yihan Liu and Yudong Cui for their very helpful discussions and suggestions. This work is supported by the National Natural Science Foundation of China (NSFC) grants 12261141691. 

\end{acknowledgments}


\bibliography{stellar_wind}

@article{amenomoriPotentialPeVatronSupernova2021,
    author = "Amenomori, M. and others",
    collaboration = "Tibet AS{\ensuremath{\gamma}}",
    title = "{Potential PeVatron supernova remnant G106.3+2.7 seen in the highest-energy gamma rays}",
    eprint = "2109.02898",
    archivePrefix = "arXiv",
    primaryClass = "astro-ph.HE",
    doi = "10.1038/s41550-020-01294-9",
    journal = "Nature Astron.",
    number = "5",
    pages = "460--464",
    year = "2021"
}

@article{apelKASCADEGrandeMeasurementsEnergy2013,
    author = "Apel, W. D. and others",
    title = "{KASCADE-Grande measurements of energy spectra for elemental groups of cosmic rays}",
    eprint = "1306.6283",
    archivePrefix = "arXiv",
    primaryClass = "astro-ph.HE",
    doi = "10.1016/j.astropartphys.2013.06.004",
    journal = "Astropart. Phys.",
    volume = "47",
    pages = "54--66",
    year = "2013"
}

@article{caoUltrahighenergyPhotons142021,
    author = "Cao, Zhen and others",
    collaboration = "LHAASO",
    title = "{Ultrahigh-energy photons up to 1.4 petaelectronvolts from 12 $\gamma$-ray Galactic sources}",
    doi = "10.1038/s41586-021-03498-z",
    journal = "Nature",
    volume = "594",
    number = "7861",
    pages = "33--36",
    year = "2021"
}

@article{choiMesaIsochronesStellar2016,
  title = {Mesa {{Isochrones}} and {{Stellar Tracks}} ({{MIST}}). {{I}}. {{Solar-scaled Models}}},
  author = {Choi, Jieun and Dotter, Aaron and Conroy, Charlie and Cantiello, Matteo and Paxton, Bill and Johnson, Benjamin D.},
  year = 2016,
  month = jun,
  journal = {The Astrophysical Journal},
  volume = {823},
  pages = {102},
  publisher = {IOP},
  issn = {0004-637X},
  doi = {10.3847/0004-637X/823/2/102},
  urldate = {2025-12-10}
}

@article{collaborationAccelerationPetaelectronvoltProtons2016,
    author = "Abramowski, A. and others",
    collaboration = "H.E.S.S.",
    title = "{Acceleration of petaelectronvolt protons in the Galactic Centre}",
    eprint = "1603.07730",
    archivePrefix = "arXiv",
    primaryClass = "astro-ph.HE",
    doi = "10.1038/nature17147",
    journal = "Nature",
    volume = "531",
    pages = "476",
    year = "2016"
}

@article{collaborationEnergySpectraElemental2009,
  title = {Energy {{Spectra}} of {{Elemental Groups}} of {{Cosmic Rays}}: {{Update}} on the {{KASCADE Unfolding Analysis}}},
  shorttitle = {Energy {{Spectra}} of {{Elemental Groups}} of {{Cosmic Rays}}},
  author = {Collaboration, {\relax KASCADE} and Apel, W. D.},
  year = 2009,
  month = mar,
  journal = {Astroparticle Physics},
  volume = {31},
  number = {2},
  eprint = {0812.0322},
  primaryclass = {astro-ph},
  pages = {86--91},
  issn = {09276505},
  doi = {10.1016/j.astropartphys.2008.11.008},
  urldate = {2025-11-20},
  archiveprefix = {arXiv}
}

@article{collaborationFirstIdentificationPrecise2025,
    author = "Cao, Zhen and others",
    collaboration = "LHAASO",
    title = "{Precise measurements of the cosmic ray proton energy spectrum in the {\textquotedblleft}knee{\textquotedblright} region}",
    eprint = "2505.14447",
    archivePrefix = "arXiv",
    primaryClass = "astro-ph.HE",
    doi = "10.1016/j.scib.2025.10.048",
    journal = "Sci. Bull.",
    volume = "70",
    pages = "4173--4180",
    year = "2025"
}

@article{collaborationPreciseMeasurementCosmic2025,
    author = "Cao, Zhen and others",
    collaboration = "LHAASO",
    title = "{Precise Measurement of the Cosmic Ray Helium Spectrum above 0.1~PeV}",
    eprint = "2511.05013",
    archivePrefix = "arXiv",
    primaryClass = "astro-ph.HE",
    doi = "10.1103/d838-49gt",
    journal = "Phys. Rev. Lett.",
    volume = "136",
    number = "12",
    pages = "121001",
    year = "2026"
}

@article{collaborationUltrahighenergyGammarayBubble2024,
    author = "Cao, Zhen and others",
    collaboration = "LHAASO",
    title = "{An ultrahigh-energy {\ensuremath{\gamma}}-ray bubble powered by a super PeVatron}",
    eprint = "2310.10100",
    archivePrefix = "arXiv",
    primaryClass = "astro-ph.HE",
    doi = "10.1016/j.scib.2023.12.040",
    journal = "Sci. Bull.",
    volume = "69",
    number = "4",
    pages = "449--457",
    year = "2024"
}

@article{collaborationUltrahighEnergyGammarayEmission2025,
  title = {Ultrahigh-{{Energy Gamma-ray Emission Associated}} with {{Black Hole-Jet Systems}}},
  author = {Collaboration, {\relax LHAASO}},
  year = 2025,
  month = nov,
  journal = {National Science Review},
  eprint = {2410.08988},
  primaryclass = {astro-ph},
  pages = {nwaf496},
  issn = {2095-5138, 2053-714X},
  doi = {10.1093/nsr/nwaf496},
  urldate = {2025-11-19},
  archiveprefix = {arXiv}
}

@article{icecubecollaborationCosmicRaySpectrum2019,
    author = "Aartsen, M. G. and others",
    collaboration = "IceCube",
    title = "{Cosmic ray spectrum and composition from PeV to EeV using 3 years of data from IceTop and IceCube}",
    eprint = "1906.04317",
    archivePrefix = "arXiv",
    primaryClass = "astro-ph.HE",
    doi = "10.1103/PhysRevD.100.082002",
    journal = "Phys. Rev. D",
    volume = "100",
    number = "8",
    pages = "082002",
    year = "2019"
}

@article{lagageMaximumEnergyCosmic1983,
  title = {The Maximum Energy of Cosmic Rays Accelerated by Supernova Shocks.},
  author = {Lagage, P. O. and Cesarsky, C. J.},
  year = 1983,
  month = sep,
  journal = {Astronomy and Astrophysics},
  volume = {125},
  pages = {249--257},
  publisher = {EDP},
  issn = {0004-6361},
  urldate = {2025-11-19}
}

@article{morlinoParticleAccelerationWinds2021,
  title = {Particle Acceleration in Winds of Star Clusters},
  author = {Morlino, G. and Blasi, P. and Peretti, E. and Cristofari, P.},
  year = 2021,
  month = jul,
  journal = {Monthly Notices of the Royal Astronomical Society},
  volume = {504},
  pages = {6096--6105},
  publisher = {OUP},
  issn = {0035-8711},
  doi = {10.1093/mnras/stab690},
  urldate = {2025-12-21}
}

@article{castorRadiationdrivenWindsStars1975,
  title = {Radiation-Driven Winds in {{Of}} Stars.},
  author = {Castor, J. I. and Abbott, D. C. and Klein, R. I.},
  year = 1975,
  month = jan,
  journal = {The Astrophysical Journal},
  volume = {195},
  pages = {157--174},
  publisher = {IOP},
  issn = {0004-637X},
  doi = {10.1086/153315},
  urldate = {2025-12-19}
}

@article{nugisMasslossRatesWolfRayet2000,
  title = {Mass-Loss Rates of {{Wolf-Rayet}} Stars as a Function of Stellar Parameters},
  author = {Nugis, T. and Lamers, H. J. G. L. M.},
  year = 2000,
  month = aug,
  journal = {Astronomy and Astrophysics},
  volume = {360},
  pages = {227--244},
  publisher = {EDP},
  issn = {0004-6361},
  urldate = {2025-12-21}
}

@article{weaverInterstellarBubblesII1977,
  title = {Interstellar Bubbles. {{II}}. {{Structure}} and Evolution.},
  author = {Weaver, R. and McCray, R. and Castor, J. and Shapiro, P. and Moore, R.},
  year = 1977,
  month = dec,
  journal = {The Astrophysical Journal},
  volume = {218},
  pages = {377--395},
  publisher = {IOP},
  issn = {0004-637X},
  doi = {10.1086/155692},
  urldate = {2025-12-21}
}

@ARTICLE{2001A&A...369..574V,
       author = {{Vink}, Jorick S. and {de Koter}, A. and {Lamers}, H.~J.~G.~L.~M.},
        title = "{Mass-loss predictions for O and B stars as a function of metallicity}",
      journal = {\aap},
         year = 2001,
        month = apr,
       volume = {369},
        pages = {574-588},
          doi = {10.1051/0004-6361:20010127},
archivePrefix = {arXiv},
       eprint = {astro-ph/0101509},
 primaryClass = {astro-ph},
       adsurl = {https://ui.adsabs.harvard.edu/abs/2001A&A...369..574V}
}

@article{malkovNonlinearTheoryDiffusive2001,
  title = {Nonlinear Theory of Diffusive Acceleration of Particles by Shock Waves},
  author = {Malkov, M. A. and Drury, L. O'C.},
  year = 2001,
  month = apr,
  journal = {Reports on Progress in Physics},
  volume = {64},
  pages = {429--481},
  publisher = {IOP},
  issn = {0034-4885},
  doi = {10.1088/0034-4885/64/4/201},
  urldate = {2026-01-22}
}

@ARTICLE{2025arXiv251106733Y,
       author = {{Yuan}, Qiang},
        title = "{Implication of multiple source populations of Galactic cosmic rays from proton and helium spectra}",
      journal = {arXiv e-prints},
         year = 2025,
        month = nov,
          eid = {arXiv:2511.06733},
        pages = {arXiv:2511.06733},
          doi = {10.48550/arXiv.2511.06733},
archivePrefix = {arXiv},
       eprint = {2511.06733},
 primaryClass = {astro-ph.HE},
       adsurl = {https://ui.adsabs.harvard.edu/abs/2025arXiv251106733Y}
}

@article{2026NatAs.tmp...79L,
    author = "Cao, Zhen and others",
    collaboration = "LHAASO",
    title = "{An extreme particle accelerator powered by PSR J1849-0001}",
    journal = {Nature Astronomy},
         year = 2026,
        month = apr,
          doi = {10.1038/s41550-026-02839-0},
archivePrefix = {arXiv},
       eprint = {2603.15537},
 primaryClass = {astro-ph.HE},
       adsurl = {https://ui.adsabs.harvard.edu/abs/2026NatAs.tmp...79L}
}

@ARTICLE{2013MNRAS.429.1156S,
       author = {{Seitenzahl}, Ivo R. and {Ciaraldi-Schoolmann}, Franco and {R{\"o}pke}, Friedrich K. and {Fink}, Michael and {Hillebrandt}, Wolfgang and {Kromer}, Markus and {Pakmor}, R{\"u}diger and {Ruiter}, Ashley J. and {Sim}, Stuart A. and {Taubenberger}, Stefan},
        title = "{Three-dimensional delayed-detonation models with nucleosynthesis for Type Ia supernovae}",
      journal = {\mnras},
         year = 2013,
        month = feb,
       volume = {429},
       number = {2},
        pages = {1156-1172},
          doi = {10.1093/mnras/sts402},
archivePrefix = {arXiv},
       eprint = {1211.3015},
 primaryClass = {astro-ph.SR},
       adsurl = {https://ui.adsabs.harvard.edu/abs/2013MNRAS.429.1156S}
}

@ARTICLE{2020SSRv..216...42B,
       author = {{Bykov}, Andrei M. and {Marcowith}, Alexandre and {Amato}, Elena and {Kalyashova}, Maria E. and {Kruijssen}, J.~M. Diederik and {Waxman}, Eli},
        title = "{High-Energy Particles and Radiation in Star-Forming Regions}",
      journal = {\ssr},
         year = 2020,
        month = apr,
       volume = {216},
       number = {3},
          eid = {42},
        pages = {42},
          doi = {10.1007/s11214-020-00663-0},
archivePrefix = {arXiv},
       eprint = {2003.11534},
 primaryClass = {astro-ph.HE},
       adsurl = {https://ui.adsabs.harvard.edu/abs/2020SSRv..216...42B}
}

@ARTICLE{1983SSRv...36..173C,
       author = {{Cesarsky}, C.~J. and {Montmerle}, T.},
        title = "{Gamma-Rays from Active Regions in the Galaxy - the Possible Contribution of Stellar Winds}",
      journal = {\ssr},
         year = 1983,
        month = oct,
       volume = {36},
       number = {2},
        pages = {173-193},
          doi = {10.1007/BF00167503},
       adsurl = {https://ui.adsabs.harvard.edu/abs/1983SSRv...36..173C}
}

@ARTICLE{2020MNRAS.493.3159G,
       author = {{Gupta}, Siddhartha and {Nath}, Biman B. and {Sharma}, Prateek and {Eichler}, David},
        title = "{Realistic modelling of wind and supernovae shocks in star clusters: addressing $^{22}$Ne/$^{20}$Ne and other problems in Galactic cosmic rays}",
      journal = {\mnras},
         year = 2020,
        month = apr,
       volume = {493},
       number = {3},
        pages = {3159-3177},
          doi = {10.1093/mnras/staa286},
archivePrefix = {arXiv},
       eprint = {1910.10168},
 primaryClass = {astro-ph.HE},
       adsurl = {https://ui.adsabs.harvard.edu/abs/2020MNRAS.493.3159G}
}

@ARTICLE{2021MNRAS.504.6096M,
       author = {{Morlino}, G. and {Blasi}, P. and {Peretti}, E. and {Cristofari}, P.},
        title = "{Particle acceleration in winds of star clusters}",
      journal = {\mnras},
         year = 2021,
        month = jul,
       volume = {504},
       number = {4},
        pages = {6096-6105},
          doi = {10.1093/mnras/stab690},
archivePrefix = {arXiv},
       eprint = {2102.09217},
 primaryClass = {astro-ph.HE},
       adsurl = {https://ui.adsabs.harvard.edu/abs/2021MNRAS.504.6096M}
}

@ARTICLE{2023MNRAS.519..136V,
       author = {{Vieu}, T. and {Reville}, B.},
        title = "{Massive star cluster origin for the galactic cosmic ray population at very-high energies}",
      journal = {\mnras},
         year = 2023,
        month = feb,
       volume = {519},
       number = {1},
        pages = {136-147},
          doi = {10.1093/mnras/stac3469},
archivePrefix = {arXiv},
       eprint = {2211.11625},
 primaryClass = {astro-ph.HE},
       adsurl = {https://ui.adsabs.harvard.edu/abs/2023MNRAS.519..136V}
}

@ARTICLE{2019NatAs...3..561A,
       author = {{Aharonian}, Felix and {Yang}, Ruizhi and {de O{\~n}a Wilhelmi}, Emma},
        title = "{Massive stars as major factories of Galactic cosmic rays}",
      journal = {Nature Astronomy},
         year = 2019,
        month = mar,
       volume = {3},
        pages = {561-567},
          doi = {10.1038/s41550-019-0724-0},
archivePrefix = {arXiv},
       eprint = {1804.02331},
 primaryClass = {astro-ph.HE},
       adsurl = {https://ui.adsabs.harvard.edu/abs/2019NatAs...3..561A}
}

@article{Peters:1961mxb,
    author = "Peters, B.",
    title = "{Primary cosmic radiation and extensive air showers}",
    doi = "10.1007/bf02783106",
    journal = "Nuovo Cim.",
    volume = "22",
    number = "4",
    pages = "800--819",
    year = "1961"
}

@article{DAMPE:2025opn,
    author = "Alemanno, Francesca and others",
    collaboration = "DAMPE",
    title = "{Charge-dependent spectral softenings of primary cosmic rays below the knee}",
    eprint = "2511.05409",
    archivePrefix = "arXiv",
    primaryClass = "astro-ph.HE",
    doi = "10.1038/s41586-026-10472-0",
    journal = "Nature",
    volume = "653",
    number = "8113",
    pages = "52--55",
    year = "2026"
}

@ARTICLE{2001MNRAS.322..231K,
       author = {{Kroupa}, Pavel},
        title = "{On the variation of the initial mass function}",
      journal = {\mnras},
         year = 2001,
        month = apr,
       volume = {322},
       number = {2},
        pages = {231-246},
          doi = {10.1046/j.1365-8711.2001.04022.x},
archivePrefix = {arXiv},
       eprint = {astro-ph/0009005},
 primaryClass = {astro-ph},
       adsurl = {https://ui.adsabs.harvard.edu/abs/2001MNRAS.322..231K}
}

@ARTICLE{2019ARA&A..57..227K,
       author = {{Krumholz}, Mark R. and {McKee}, Christopher F. and {Bland-Hawthorn}, Joss},
        title = "{Star Clusters Across Cosmic Time}",
      journal = {\araa},
         year = 2019,
        month = aug,
       volume = {57},
        pages = {227-303},
          doi = {10.1146/annurev-astro-091918-104430},
archivePrefix = {arXiv},
       eprint = {1812.01615},
 primaryClass = {astro-ph.GA},
       adsurl = {https://ui.adsabs.harvard.edu/abs/2019ARA&A..57..227K}
}

@ARTICLE{2022ApJ...941..162E,
       author = {{Elia}, D. and {Molinari}, S. and {Schisano}, E. and {Soler}, J.~D. and {Merello}, M. and {Russeil}, D. and {Veneziani}, M. and {Zavagno}, A. and {Noriega-Crespo}, A. and {Olmi}, L. and {Benedettini}, M. and {Hennebelle}, P. and {Klessen}, R.~S. and {Leurini}, S. and {Paladini}, R. and {Pezzuto}, S. and {Traficante}, A. and {Eden}, D.~J. and {Martin}, P.~G. and {Sormani}, M. and {Coletta}, A. and {Colman}, T. and {Plume}, R. and {Maruccia}, Y. and {Mininni}, C. and {Liu}, S.~J.},
        title = "{The Star Formation Rate of the Milky Way as Seen by Herschel}",
      journal = {\apj},
         year = 2022,
        month = dec,
       volume = {941},
       number = {2},
          eid = {162},
        pages = {162},
          doi = {10.3847/1538-4357/aca27d},
archivePrefix = {arXiv},
       eprint = {2211.05573},
 primaryClass = {astro-ph.GA},
       adsurl = {https://ui.adsabs.harvard.edu/abs/2022ApJ...941..162E}
}

@ARTICLE{2010ApJ...725.1886L,
       author = {{Lamb}, J.~B. and {Oey}, M.~S. and {Werk}, J.~K. and {Ingleby}, L.~D.},
        title = "{The Sparsest Clusters with O Stars}",
      journal = {\apj},
         year = 2010,
        month = dec,
       volume = {725},
       number = {2},
        pages = {1886-1902},
          doi = {10.1088/0004-637X/725/2/1886},
archivePrefix = {arXiv},
       eprint = {1010.5273},
 primaryClass = {astro-ph.GA},
       adsurl = {https://ui.adsabs.harvard.edu/abs/2010ApJ...725.1886L}
}

@ARTICLE{2022ApJS..262...30P,
       author = {{Porter}, T.~A. and {J{\'o}hannesson}, G. and {Moskalenko}, I.~V.},
        title = "{The GALPROP Cosmic-ray Propagation and Nonthermal Emissions Framework: Release v57}",
      journal = {\apjs},
         year = 2022,
        month = sep,
       volume = {262},
       number = {1},
          eid = {30},
        pages = {30},
          doi = {10.3847/1538-4365/ac80f6},
archivePrefix = {arXiv},
       eprint = {2112.12745},
 primaryClass = {astro-ph.HE},
       adsurl = {https://ui.adsabs.harvard.edu/abs/2022ApJS..262...30P}
}

@article{2023PhRvL.130u1002A,
    author = "Aguilar, M. and others",
    collaboration = "AMS",
    title = "{Properties of Cosmic-Ray Sulfur and Determination of the Composition of Primary Cosmic-Ray Carbon, Neon, Magnesium, and Sulfur: Ten-Year Results from the Alpha Magnetic Spectrometer}",
    doi = "10.1103/PhysRevLett.130.211002",
    journal = "Phys. Rev. Lett.",
    volume = "130",
    number = "21",
    pages = "211002",
    year = "2023"
}

@ARTICLE{2019AdSpR..64.2546G,
       author = {{Grebenyuk}, V. and {Karmanov}, D. and {Kovalev}, I. and {Kudryashov}, I. and {Kurganov}, A. and {Panov}, A. and {Podorozhny}, D. and {Tkachenko}, A. and {Tkachev}, L. and {Turundaevskiy}, A. and {Vasiliev}, O. and {Voronin}, A.},
        title = "{Energy spectra of abundant cosmic-ray nuclei in the NUCLEON experiment}",
      journal = {Advances in Space Research},
         year = 2019,
        month = dec,
       volume = {64},
       number = {12},
        pages = {2546-2558},
          doi = {10.1016/j.asr.2019.10.004},
       adsurl = {https://ui.adsabs.harvard.edu/abs/2019AdSpR..64.2546G}
}

@article{2021PhR...894....1A,
    author = "Aguilar, M. and others",
    collaboration = "AMS",
    title = "{The Alpha Magnetic Spectrometer (AMS) on the international space station: Part II {\textemdash} Results from the first seven years}",
    doi = "10.1016/j.physrep.2020.09.003",
    journal = "Phys. Rept.",
    volume = "894",
    pages = "1--116",
    year = "2021"
}

@article{yuanPropagationCosmicRays2017,
  title = {Propagation of Cosmic Rays in the {{AMS-02}} Era},
  author = {Yuan, Qiang and Lin, Su-Jie and Fang, Kun and Bi, Xiao-Jun},
  year = 2017,
  journal = {Physical Review},
  volume = {D95},
  number = {8},
  eprint = {1701.06149},
  primaryclass = {astro-ph.HE},
  pages = {083007},
  doi = {10.1103/PhysRevD.95.083007},
  archiveprefix = {arXiv}
}

@article{maInterpretationsCosmicRay2023,
  title = {Interpretations of the Cosmic Ray Secondary-to-Primary Ratios Measured by {{DAMPE}}},
  author = {Ma, Peng-Xiong and Xu, Zhi-Hui and Yuan, Qiang and Bi, Xiao-Jun and Fan, Yi-Zhong and Moskalenko, Igor V. and Yue, Chuan},
  year = 2023,
  month = aug,
  journal = {Frontiers of Physics},
  volume = {18},
  number = {4},
  eprint = {2210.09205},
  primaryclass = {astro-ph},
  pages = {44301},
  issn = {2095-0462, 2095-0470},
  doi = {10.1007/s11467-023-1257-7},
  urldate = {2025-11-02},
  archiveprefix = {arXiv}
}

@article{gleesonSolarModulationGalactic1968,
  title = {Solar {{Modulation}} of {{Galactic Cosmic Rays}}},
  author = {Gleeson, L. J. and Axford, W. I.},
  year = 1968,
  journal = {Astrophys. J.},
  volume = {154},
  pages = {1011},
  doi = {10.1086/149822}
}

@ARTICLE{2024NatAs...8..530P,
       author = {{Peron}, Giada and {Casanova}, Sabrina and {Gabici}, Stefano and {Baghmanyan}, Vardan and {Aharonian}, Felix},
        title = "{The contribution of winds from star clusters to the Galactic cosmic-ray population}",
      journal = {Nature Astronomy},
         year = 2024,
        month = apr,
       volume = {8},
        pages = {530-537},
          doi = {10.1038/s41550-023-02168-6},
archivePrefix = {arXiv},
       eprint = {2407.07509},
 primaryClass = {astro-ph.HE},
       adsurl = {https://ui.adsabs.harvard.edu/abs/2024NatAs...8..530P}
}

@article{HESS:2024xux,
    author = "Aharonian, F. and others",
    collaboration = "H.E.S.S.",
    title = "{Very-high-energy $\gamma$-ray emission from young massive star clusters in the Large Magellanic Cloud}",
    eprint = "2407.16219",
    archivePrefix = "arXiv",
    primaryClass = "astro-ph.HE",
    doi = "10.3847/2041-8213/ad5e67",
    journal = "Astrophys. J. Lett.",
    volume = "970",
    number = "1",
    pages = "L21",
    year = "2024"
}

@ARTICLE{2013ApJ...764...21C,
       author = {{Chieffi}, Alessandro and {Limongi}, Marco},
        title = "{Pre-supernova Evolution of Rotating Solar Metallicity Stars in the Mass Range 13-120 M $_{☉}$ and their Explosive Yields}",
      journal = {\apj},
         year = 2013,
        month = feb,
       volume = {764},
       number = {1},
          eid = {21},
        pages = {21},
          doi = {10.1088/0004-637X/764/1/21},
       adsurl = {https://ui.adsabs.harvard.edu/abs/2013ApJ...764...21C}
}

@ARTICLE{2024ARA&A..62...21M,
       author = {{Marchant}, Pablo and {Bodensteiner}, Julia},
        title = "{The Evolution of Massive Binary Stars}",
      journal = {\araa},
         year = 2024,
        month = sep,
       volume = {62},
       number = {1},
        pages = {21-61},
          doi = {10.1146/annurev-astro-052722-105936},
archivePrefix = {arXiv},
       eprint = {2311.01865},
 primaryClass = {astro-ph.SR},
       adsurl = {https://ui.adsabs.harvard.edu/abs/2024ARA&A..62...21M}
}

@ARTICLE{2020MNRAS.499..873S,
       author = {{Sander}, Andreas A.~C. and {Vink}, Jorick S.},
        title = "{On the nature of massive helium star winds and Wolf-Rayet-type mass-loss}",
      journal = {\mnras},
         year = 2020,
        month = nov,
       volume = {499},
       number = {1},
        pages = {873-892},
          doi = {10.1093/mnras/staa2712},
archivePrefix = {arXiv},
       eprint = {2009.01849},
 primaryClass = {astro-ph.SR},
       adsurl = {https://ui.adsabs.harvard.edu/abs/2020MNRAS.499..873S}
}

@ARTICLE{2026arXiv260321665E,
       author = {{Espinosa Castro}, Luis E. and {Murase}, Kotha and {Rizza}, Carlo and {Villante}, Francesco L. and {Vecchiotti}, Vittoria and {Pagliaroli}, Giulia},
        title = "{Multimessenger Concordance for the Cygnus Region as the Source of the Cosmic-Ray Knee}",
      journal = {arXiv e-prints},
         year = 2026,
        month = mar,
          eid = {arXiv:2603.21665},
        pages = {arXiv:2603.21665},
          doi = {10.48550/arXiv.2603.21665},
archivePrefix = {arXiv},
       eprint = {2603.21665},
 primaryClass = {astro-ph.HE},
       adsurl = {https://ui.adsabs.harvard.edu/abs/2026arXiv260321665E}
}

@ARTICLE{2006MNRAS.365.1333W,
       author = {{Weidner}, Carsten and {Kroupa}, Pavel},
        title = "{The maximum stellar mass, star-cluster formation and composite stellar populations}",
      journal = {\mnras},
         year = 2006,
        month = feb,
       volume = {365},
       number = {4},
        pages = {1333-1347},
          doi = {10.1111/j.1365-2966.2005.09824.x},
archivePrefix = {arXiv},
       eprint = {astro-ph/0511331},
 primaryClass = {astro-ph},
       adsurl = {https://ui.adsabs.harvard.edu/abs/2006MNRAS.365.1333W}
}

@article{Wright:2010cc,
    author = "Wright, Nicholas J. and Drake, Jeremy J. and Drew, Janet E. and Vink, Jorick S.",
    title = "{The Massive Star Forming Region Cygnus OB2. II. Integrated Stellar Properties and the Star Formation History}",
    eprint = "1003.2463",
    archivePrefix = "arXiv",
    primaryClass = "astro-ph.SR",
    doi = "10.1088/0004-637X/713/2/871",
    journal = "Astrophys. J.",
    volume = "713",
    pages = "871--882",
    year = "2010"
}

@article{portegies_zwart_young_2010,
    author = "Portegies Zwart, Simon and McMillan, Steve and Gieles, Mark",
    title = "{Young massive star clusters}",
    eprint = "1002.1961",
    archivePrefix = "arXiv",
    primaryClass = "astro-ph.GA",
    doi = "10.1146/annurev-astro-081309-130834",
    journal = "Ann. Rev. Astron. Astrophys.",
    volume = "48",
    pages = "431--493",
    year = "2010"
}

@article{Bohm,
    author = "Drury, L. Oc.",
    title = "{An introduction to the theory of diffusive shock acceleration of energetic particles in tenuous plasmas}",
    doi = "10.1088/0034-4885/46/8/002",
    journal = "Rept. Prog. Phys.",
    volume = "46",
    pages = "973--1027",
    year = "1983"
}

@article{Kolmogorov,
    author = {Kolmogorov, Andrei Nikolaevich and Levin, V. and Hunt, Julian Charles Roland and Phillips, Owen Martin and Williams, David},
    title = {The local structure of turbulence in incompressible viscous fluid for very large Reynolds numbers},
    journal = {Proceedings of the Royal Society of London. Series A: Mathematical and Physical Sciences},
    volume = {434},
    number = {1890},
    pages = {9-13},
    year = {1991},
    month = {07},
    issn = {0962-8444},
    doi = {10.1098/rspa.1991.0075},
    url = {https://doi.org/10.1098/rspa.1991.0075},
    eprint = {https://royalsocietypublishing.org/rspa/article-pdf/434/1890/9/68114/rspa.1991.0075.pdf},
}

@article{Kraichnan:1965zz,
    author = "Kraichnan, Robert H.",
    title = "{Inertial-Range Spectrum of Hydromagnetic Turbulence}",
    doi = "10.1063/1.1761412",
    journal = "Phys. Fluids",
    volume = "8",
    pages = "1385--1387",
    year = "1965"
}

@article{IceCube-Gen2:2021tmd,
    author = "Omeliukh, Anastasiia and others",
    collaboration = "IceCube-Gen2",
    title = "{Optimization of the optical array geometry for IceCube-Gen2}",
    eprint = "2107.08527",
    archivePrefix = "arXiv",
    primaryClass = "astro-ph.HE",
    reportNumber = "PoS-ICRC2021-1184",
    doi = "10.22323/1.395.1184",
    journal = "PoS",
    volume = "ICRC2021",
    pages = "1184",
    year = "2021"
}

@article{KM3NeT:2018wnd,
    author = "Aiello, S. and others",
    collaboration = "KM3NeT",
    title = "{Sensitivity of the KM3NeT/ARCA neutrino telescope to point-like neutrino sources}",
    eprint = "1810.08499",
    archivePrefix = "arXiv",
    primaryClass = "astro-ph.HE",
    doi = "10.1016/j.astropartphys.2019.04.002",
    journal = "Astropart. Phys.",
    volume = "111",
    pages = "100--110",
    year = "2019"
}

@article{TRIDENT:proposal,
    author = "Ye, Z. P. and others",
    title = "{A multi-cubic-kilometre neutrino telescope in the western Pacific Ocean}",
    doi = "10.1038/s41550-023-02087-6",
    journal = "Nature Astron.",
    volume = "7",
    number = "12",
    pages = "1497--1505",
    year = "2023"
}

@article{HUNT:2023mzt,
    author = "Huang, Tian-Qi and Cao, Zhen and Chen, Mingjun and Liu, Jiali and Wang, Zike and You, Xiaohao and Qi, Ying",
    title = "{Proposal for the High Energy Neutrino Telescope}",
    doi = "10.22323/1.444.1080",
    journal = "PoS",
    volume = "ICRC2023",
    pages = "1080",
    year = "2023"
}

\end{document}